\documentclass[10pt]{article}

\usepackage{amsfonts,amssymb,amsmath,mathrsfs,latexsym,bbm,dsfont,amsthm}
\usepackage{color}
\usepackage{graphicx}
\usepackage{pstricks,pstricks-add}
\usepackage{pst-plot}
\usepackage[scriptsize,nooneline,hang]{caption}
\usepackage[hang,nooneline,scriptsize]{subfigure}
\usepackage{rotating}
\usepackage[all]{xy}
\usepackage[abs]{overpic}

\newcommand{\half}{\frac{1}{2}}
\newcommand{\tr}{\tilde{r}}
\newcommand{\tw}{\tilde{w}}
\newcommand{\tlt}{\tilde{t}}
\newcommand{\tL}{\tilde{L}}
\newcommand{\tK}{\tilde{K}}
\newcommand{\ta}{\tilde{a}}
\newcommand{\trho}{\tilde{\rho}}
\newcommand{\dd}{{\rm d}}
\newcommand{\Dr}{\Delta_r}
\newcommand{\tDr}{\tilde{\Delta}_r}

\newcommand{\tv}{\tilde{v}}

\begin{document}

\title{Geodesic motion in the (rotating) black string spacetime}

\author{Saskia Grunau$^1$, Bhavesh Khamesra$^2$\\
$^1$ Institut f\"ur Physik, Universit\"at Oldenburg, D--26111 Oldenburg, Germany\\
$^2$ IISER (Indian Institute of Science Education and Research) Pune, Maharashtra 411008, India
}

\maketitle

\begin{abstract}
In this article we study the geodesic motion of test particles and light in the five-dimensional (rotating) black string spacetime. If a compact dimension is added to the four-dimensional Schwarzschild or Kerr spacetime, the new five-dimensional metric describes a (rotating) black string. The geodesics in the Schwarzschild and Kerr spacetime have been studied in great detail, however, when a compact dimension is added new behaviour occurs. We present the analytical solutions of the geodesic equations and discuss the possible orbits. The motion in the ordinary four-dimensional Schwarzschild  and Kerr spacetime is compared to the motion in the (rotating) black string spacetime.
\end{abstract}

\section{Introduction}

In 1916 Schwarzschild introduced the first exact solution of the vacuum Einstein equations in four dimensions \cite{Schwarzschild:1916ae}. This unique solution describes a spherically symmetric black hole. About fifty years later, in 1963, the solution of a rotating black hole was discovered by Kerr \cite{Kerr:1963ud}.

If a compact dimension is added to the Schwarzschild metric, it describes a five-dimensional black string. Analogously the Kerr spacetime turns into the spacetime of a rotating black string if a compact dimension is added.

The idea of compact dimensions came up in the 1920s after Kaluza suggested to add a fifth dimension to general relativity \cite{Kaluza:1921tu}. This was one of the first attempts to unify gravity and the electromagnetic forces. In 1926 Klein proposed a physical interpretation \cite{Klein:1926tv}: The added dimension is compact, i.e., it is curled up in itself and has a certain length, which is too small to be observed. But still the theory of Kaluza and Klein could not explain the weakness of gravity in comparison to the electromagnetic forces.

In the 60s and 70s the string theory was developed. Here the existence of higher dimensions is essential for the internal consistency of the theory, mostly they are also assumed to be compact.\\

The equations of motion for test particles in the Schwarzschild spacetime were solved in 1931 by Hagihara \cite{Hagihara:1931} in terms of the elliptic  Weierstra{\ss} $\wp$-, $\sigma$- and $\zeta$-functions. In the Taub-NUT \cite{Kagramanova:2010bk}, Reissner-Nordstr\"om \cite{Grunau:2010gd} and Myers-Perry \cite{Kagramanova:2012hw} spacetime the equations of motion were also solved in terms of the elliptic Weierstra{\ss} functions.

The integration of geodesics was advanced in the papers of Hackmann and L\"ammerzahl \cite{Hackmann:2008zza,Hackmann:2008zz}. They integrated the geodesics in the four-dimensional Schwarzschild-de Sitter space-time analytically in terms of the hyperelliptic $\theta$- and $\sigma$-functions. The mathematical problem is based on the Jacobi inversion problem which can be solved if restriced to the $\theta$-divisor. This method was also applied to find the analytical solution of the equations of motion in higher dimensional Schwarzschild, Schwarzschild-(anti)de Sitter, Reissner-Nordstr\"om and Reissner-Nordstr\"om -(anti) de Sitter spactime \cite{Hackmann:2008tu} as well as in Kerr-(anti) de Sitter spacetimes \cite{Hackmann:2010zz} and in higher dimensional Myers-Perry spacetime \cite{Enolski:2010if}. Recently, the geodesics equations in special cases were solved analytically in the singly spinning black ring spacetime \cite{Grunau:2012ai} and in the (charged) doubly spinning black ring spacetime \cite{Grunau:2012ri}.\\

The test particle motion in different spacetimes containing black strings was studied in \cite{Aliev:1988wv,Galtsov:1989ct,Chakraborty:1991mb,Ozdemir:2003km,Ozdemir:2004ne}. In \cite{Hackmann:2009rp} and \cite{Hackmann:2010ir} the analytical solutions of the equations of motion in the Schwarzschild and Kerr spacetime pierced by black string were presented. Furthermore, the geodesic motion of test particles in field theoretical cosmic string spacetimes were investigated, namely Abelian-Higgs strings \cite{Hartmann:2010rr}, two interacting Abelian-Higgs strings \cite{Hartmann:2012pj} and cosmic superstrings \cite{Hartmann:2010vp}.\\

In this paper we will consider static and rotating black strings, which can be obtained by adding an extra compact dimension to the Schwarzschild and Kerr metric. The geodesic motion in the ordinary Schwarzschild spacetime and the Kerr spacetime has already been analyzed in detail, however, when a compact dimension is added new behaviour occurs which is studied in this article. We present the analytic solutions of the geodesic equations in terms of the elliptic Weierstra{\ss} functions and discuss the corresponding orbits. In the first part the non-rotating black string is studied and compared to the Schwarzschild black hole. In the second part the rotating black string is analysed and compared to the Kerr black hole.

\section{Black string spacetime}

In this section we will discuss the geodesics in the static black string spacetime and present analytical solutions of the equations of motion.

\subsection{The geodesic equations}

The well known Schwarzschild metric is the unique static spherically symmetric solution to the vacuum Einstein equations in four dimensions. If an extra compact spatial dimension $w$ is added, the metric takes the form
\begin{equation}
 {\rm d}s^2=-\left(1-\frac{2M}{r}\right){\rm d}t^2 + \left(\frac{1}{1-\frac{2M}{r}}\right){\rm d}r^2 + r^2({\rm d}\vartheta^2 + \sin^2\vartheta{\rm d}\varphi^2) +{\rm d}w^2 \, ,
\end{equation}
where $M$ is proportional to the mass of the black hole. This solution describes a neutral uniform black string. As usual the singularity is located at $r=0$. The horizon is located at $r=2M$ and covers the extra dimension. The gravitational constant and the velocity of light are set to 1.

The Hamilton-Jacobi equation
\begin{equation}
 \frac{\partial S}{\partial \tau} + \half g^{\mu\nu} \frac{\partial S}{\partial x^\mu}\frac{\partial S}{\partial x^\nu}=0
\label{eqn:ham-jac}
\end{equation}
has a solution of the form
\begin{equation}
 S=\half\delta \tau -Et+L\varphi + Jw + S_r(r) \, .
\end{equation}
Here $\tau$ is an affine parameter along the geodesic. The parameter $\delta$ is equal to $1$ for particles and equal to $0$ for light. $E$, $L$ and $J$ are the conserved momenta; $E$ is the energy, $L$ denotes the angular momentum and $J$ is a new constant of motion according to the compact dimension $w$. We set $\vartheta=\frac{\pi}{2}$ since the orbits lie in a plane if plotted in Cartesian $x$-$y$-$z$ coordinates ($x=r\cos\varphi\sin\vartheta$, $y=r\sin\varphi\sin\vartheta$, $z=r\cos\vartheta$) due to the spherical symmetry of the original Schwarzschild metric.

For convenience, we introduce dimensionless quantities ($r_{\rm S}=2M$)
\begin{equation}
\tr=\frac{r}{r_{\rm S}} \ , \,\, \tw=\frac{w}{r_{\rm S}} \ , \,\, \tlt=\frac{t}{r_{\rm S}} \ , \,\, \tilde{\tau}=\frac{\tau}{r_{\rm S}} \ , \,\, \tL=\frac{L}{r_{\rm S}} \ .
\end{equation}
In these dimensionless quantities the horizon is at $\tr_H=1$.

The Hamilton-Jacobi equation \eqref{eqn:ham-jac} separates and yields a differential equation for each coordinate
\begin{eqnarray}
 \left(\frac{\dd \tr}{\dd \gamma}\right) ^2 &=& R \label{eqn:r-equation}\\
 \frac{\dd \varphi}{\dd \gamma} &=& \tL \label{eqn:phi-equation}\\
 \frac{\dd \tw}{\dd \gamma} &=& J\tr ^2 \label{eqn:w-equation} \\
 \frac{\dd \tlt}{\dd \gamma} &=& \frac{E\tr^3}{\tr-1}  \label{eqn:t-equation}
\end{eqnarray}
with the polynomial
\begin{equation}
 R=E^2\tr^4-(\delta + J^2)(\tr^4-\tr^3)-\tL^2(\tr^2-\tr) \, .
\end{equation}
We also used the Mino time $\gamma$ as $\tr^2\dd\gamma = \dd\tilde{\tau}$ \cite{Mino:2003yg}. The new constant of motion $J$ due to the compact dimension is only present in the $\tr$-equation \eqref{eqn:r-equation} and of course in the $\tw$-equation \eqref{eqn:w-equation}. The other equations are the same as in the original Schwarzschild spacetime without the compact dimension.

\subsection{Classification of the geodesics}

Before discussing the $\tr$-motion in detail, we introduce a list of all possible orbits:
\begin{enumerate}
 \item \textit{Terminating orbit} (TO) with ranges $\tr \in [0, \infty)$ or $\tr \in [0, r_1]$ with $r_1\geq\tr_H$. The TOs end in the singularity at $\tr=0$.
 \item \textit{Escape orbit} (EO) with range $\tr \in [r_1, \infty)$ with $r_1>\tr_H$.
 \item \textit{Bound orbit} (BO) with range $\tr \in [r_1, r_2]$ with $r_1 < r_2$ and $r_1,r_2>\tr_H$.\\
\end{enumerate}

The right hand side of the differential equation \eqref{eqn:r-equation} has the form $R=\sum _{i=1}^4 a_i\tr^i$ with the coefficients
\begin{eqnarray}
 a_4 &=& E^2-\delta-J^2\\
 a_3 &=& \delta+J^2 \\
 a_2 &=& -\tL^2 \\
 a_1 &=& \tL^2 \, .
\end{eqnarray}
$R\geq 0$ is required in order to obtain real values for $\tr$. The regions for which $R\geq 0$  are bounded by the zeros of $R$. The number of zeros can be determined with the help of the rule of Descartes. One can distinguish four cases:
\begin{enumerate}
 \item If $E^2\geq\delta+J^2$ and $\tL^2>0$, then two or no zeros are possible.
 \item If $E^2\geq\delta+J^2$ and $\tL=0$, no zeros are possible.
 \item If $E^2<\delta+J^2$ and $\tL^2>0$, then one or three zeros are possible.
 \item If $E^2<\delta+J^2$ and $\tL=0$, one zero is possible.
\end{enumerate}
Since $E$ is real and therefore $E^2\geq0$, $\delta+J^2\geq 0$ is required. So if $J=0$ (where the equations of motion reduce to the original Schwarzschild case without the compact dimension) the cases 3. and 4. are not possible for light ($\delta=0$). But if $J\neq0$ there are 3 possible zeros of $R$ for $\delta=0$, that means in the Schwarzschild spacetime with an extra compact dimension we have the possibility of stable bound orbits for light, which are not possible in the original Schwarzschild spacetime.\\

\begin{figure}[h]
 \centering
 \subfigure[$\delta=1$, $L=\sqrt{2}$ and $J=1.5$: Examples of the orbit types A, B and B$_0$.]{
   \includegraphics[width=6cm]{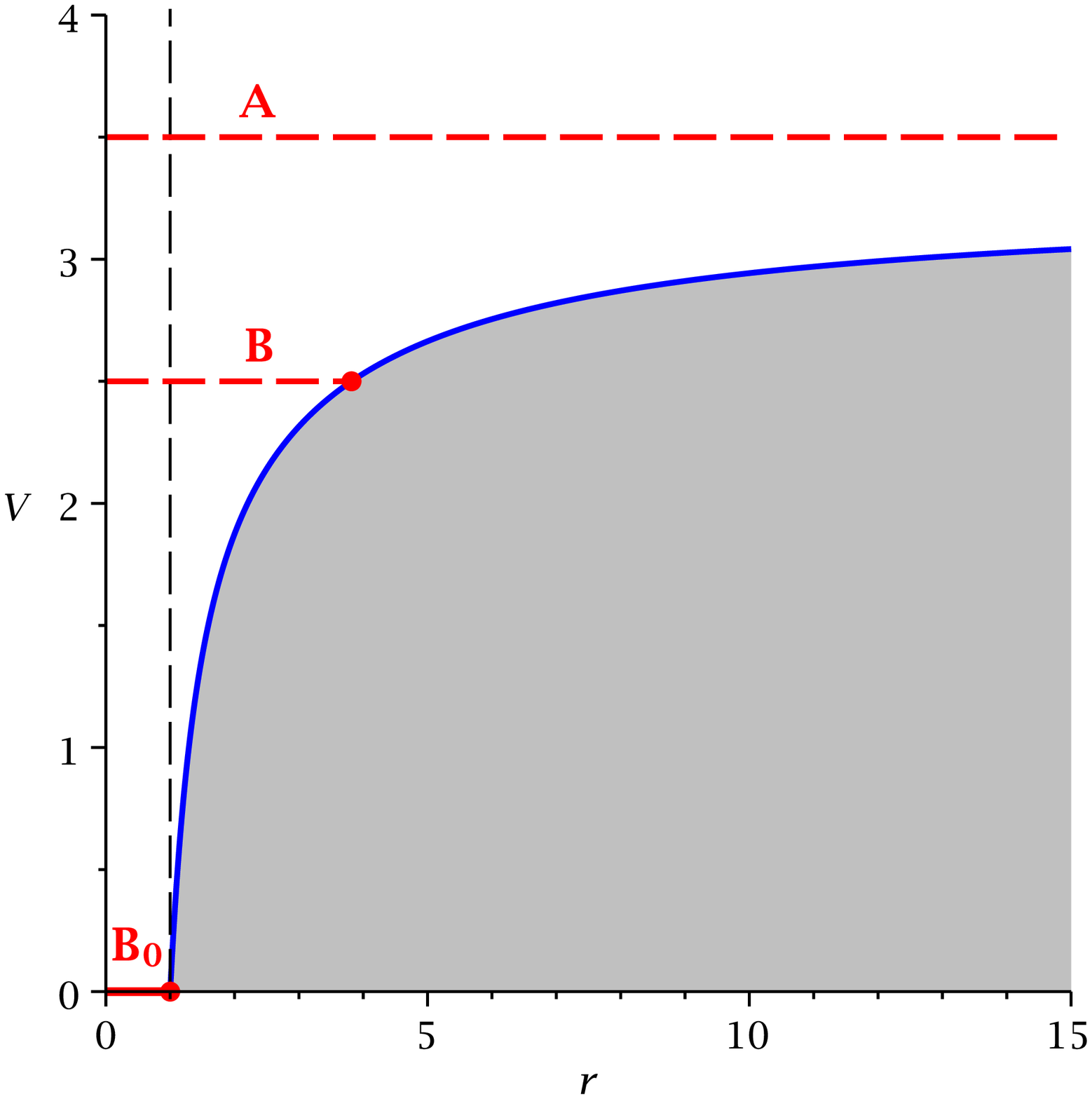}
 }
 \subfigure[$\delta=0$, $L=\sqrt{5}$ and $J=1$: Examples of the orbit types C and D. In the case of type D, here a stable bound orbit for light is possible.]{
   \includegraphics[width=6cm]{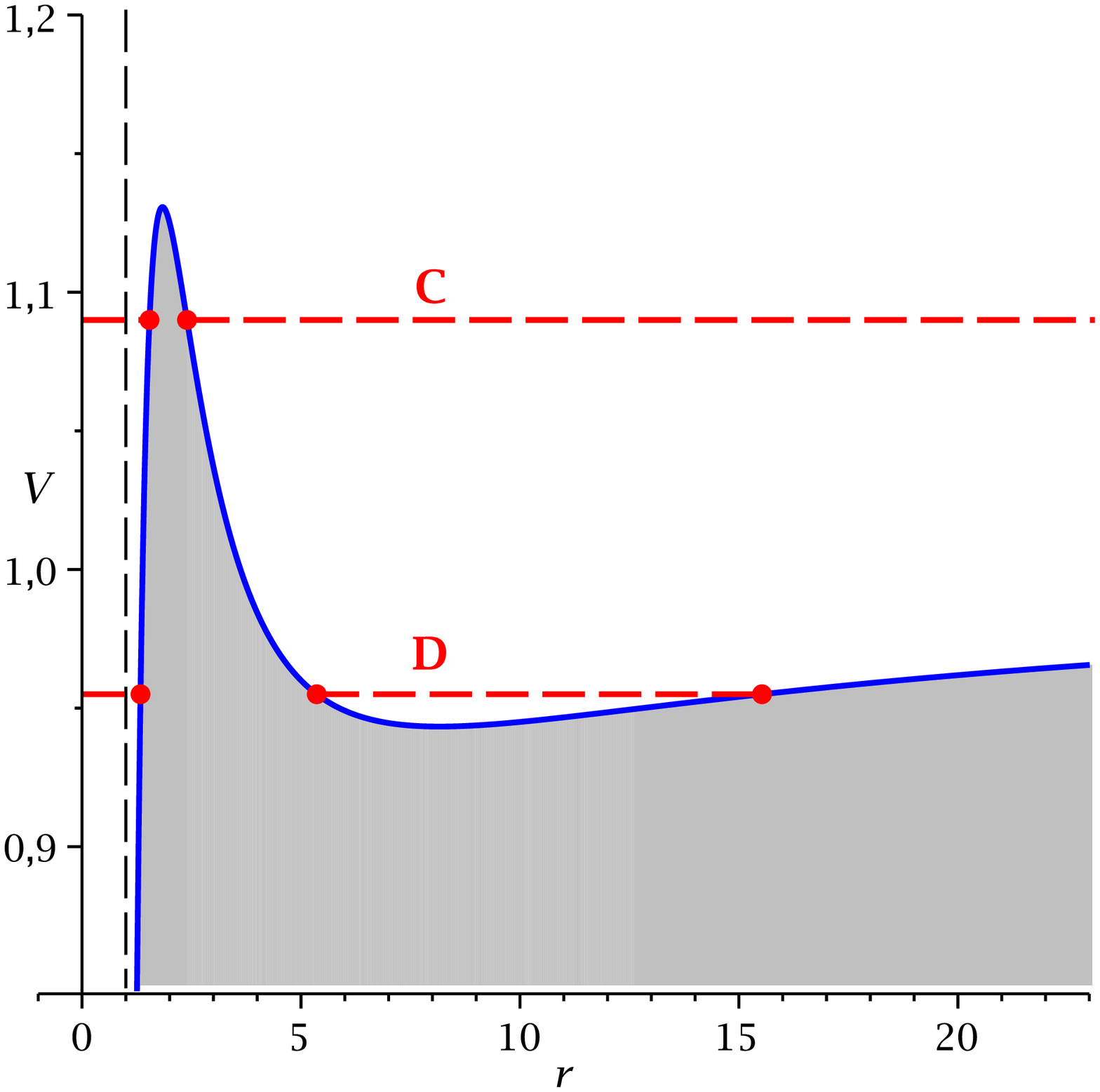}
 }
 \caption{Examples of the effective potential in the black string spacetime. The grey area is a forbidden zone, where no motion is possible. Horizontal red dashed lines represent energies and red points mark the turning points. The horizon is marked by a vertical black dashed line.}
 \label{pic:potential}
\end{figure}

We define an effective potential $V$ from the equation \eqref{eqn:r-equation} by
\begin{equation}
 \left(\frac{\dd\tr}{\dd\gamma}\right)^2=\tr^4(E^2-V)\, ,
\label{eqn:turningpoints}
\end{equation}
thus
\begin{equation}
 V=\left(1-\frac{1}{\tr}\right)\left(\delta+J^2+\frac{\tL^2}{\tr^2}\right) \, .
\end{equation}
In the limit $\tr\rightarrow\infty$ the effective potential $V$ converges to $\delta+J^2$. If $\tr\rightarrow 0$, we have $V\rightarrow -\infty$.
$\left(\frac{\dd\tr}{\dd\gamma}\right)^2=0$ determines the turning points of an orbit. Depending on the number of turning points and the shape of the effective potential four different types of orbits are possible:
\begin{itemize}
 \item Type A: no zeros. Only TOs are possible.
 \item Type B: one zero. Only TOs are possible. The turning point can coincide with the event horizon.
 \item Type C: two zeros. TOs and EOs are possible. If the energy $E$ coincides with the maximum of the effective potential then there is a single zero, which corresponds to an unstable circular bound orbit.
 \item Type D: three zeros. TOs and BOs are possible.

If the energy $E$ coincides with the minimum of the effective potential then there are two zeros. TOs and circular BOs are possible.
\end{itemize}
Table \ref{tab:type-orbits} shows an overview of the types of orbits. Some examples for energies corresponding to certain orbits in the effective potential can be seen in figure \ref{pic:potential}.

\begin{table}[h]
\begin{center}
\begin{tabular}{|lcll|}\hline
type &  zeros  & range of $\tr$ & orbit \\
\hline\hline
A & 0 & 
\begin{pspicture}(-2.5,-0.2)(3,0.2)
\psline[linewidth=0.5pt]{|->}(-2.5,0)(3,0)
\psline[linewidth=0.5pt,doubleline=true](-1,-0.2)(-1,0.2)
\psline[linewidth=1.2pt]{-}(-2.5,0)(3,0)
\end{pspicture}
  & TO
\\  \hline
B  & 1 &
\begin{pspicture}(-2.5,-0.2)(3,0.2)
\psline[linewidth=0.5pt]{|->}(-2.5,0)(3,0)
\psline[linewidth=0.5pt,doubleline=true](-1,-0.2)(-1,0.2)
\psline[linewidth=1.2pt]{-*}(-2.5,0)(0,0)
\end{pspicture}
& TO 
\\ 
B$_0$ &  & 
\begin{pspicture}(-2.5,-0.2)(3,0.2)
\psline[linewidth=0.5pt]{|->}(-2.5,0)(3,0)
\psline[linewidth=0.5pt,doubleline=true](-1,-0.2)(-1,0.2)
\psline[linewidth=1.2pt]{-*}(-2.5,0)(-1,0)
\end{pspicture}
  & TO 
\\ \hline
C & 2 & 
\begin{pspicture}(-2.5,-0.2)(3,0.2)
\psline[linewidth=0.5pt]{|->}(-2.5,0)(3,0)
\psline[linewidth=0.5pt,doubleline=true](-1,-0.2)(-1,0.2)
\psline[linewidth=1.2pt]{-*}(-2.5,0)(0,0)
\psline[linewidth=1.2pt]{*-}(1.0,0)(3,0)
\end{pspicture}
  & TO, EO
\\ \hline
D & 3 & 
\begin{pspicture}(-2.5,-0.2)(3,0.2)
\psline[linewidth=0.5pt]{|->}(-2.5,0)(3,0)
\psline[linewidth=0.5pt,doubleline=true](-1,-0.2)(-1,0.2)
\psline[linewidth=1.2pt]{-*}(-2.5,0)(0,0)
\psline[linewidth=1.2pt]{*-*}(1,0)(2,0)
\end{pspicture}
& TO, BO 
\\ \hline\hline
\end{tabular}
\caption{Types of orbits of light and particles in the black string spacetime. The thick lines represent the range of the orbits. The turning points are shown by thick dots. The horizon is indicated by a vertical double line.}
\label{tab:type-orbits}
\end{center}
\end{table}

\subsection{Solution of the geodesic equations}

In this section we present the analytical solution of the equations of motion \eqref{eqn:r-equation}-\eqref{eqn:t-equation}.

\subsubsection{The $\tr$-equation}

The polynomial $R=\sum _{i=1}^4 a_i\tr^i$ can be reduced to cubic order by the substitution $\tr=\pm\frac{1}{x}$: $R'= \sum _{i=0}^3 b_i x^i$. A further substitution $x=\frac{1}{b_3}\left( 4y-\frac{b_2}{3}\right)$ transforms $R'$ into the Weierstra{\ss} form so that equation \eqref{eqn:r-equation} turns into
\begin{equation}
\left(\frac{dy}{d\gamma}\right)^2=4y^3-g_2y-g_3= P_3(y) \, ,
\label{eqn:weierstrass}
\end{equation}
where 
\begin{equation}
g_2=\frac{b_2^2}{12} - \frac{b_1b_3}{4} \, , \qquad  g_3=\frac{b_1b_2b_3}{48} - \frac{b_0b_3^2}{16}-\frac{b_2^3}{216} \ .
\end{equation}
The differential equation \eqref{eqn:weierstrass} is of elliptic type and is solved by the Weierstra{\ss} $\wp$-function \cite{Markushevich:1967}
\begin{equation}
y(\gamma) = \wp\left(\gamma - \gamma'_{\rm in}; g_2, g_3\right) \ ,
\end{equation}
where $\gamma'_{\rm in}=\gamma_{\rm in}+\int^\infty_{y_{\rm in}}{\frac{dy}{\sqrt{4y^3-g_2y-g_3}}}$
with $y_{\rm in}=\pm\frac{b_3}{4\tr_{\rm in}} + \frac{b_2}{12}$.
Then the solution of \eqref{eqn:r-equation} acquires the form
\begin{equation}
\tr=\pm \frac{b_3}{4 \wp\left(\gamma - \gamma'_{\rm in}; g_2, g_3\right) - \frac{b_2}{3}} \ .
\end{equation}

\subsubsection{The $\varphi$-equation}

Equation \eqref{eqn:phi-equation} has the trivial solution
\begin{equation}
 \varphi(\gamma)=\tL (\gamma - \gamma_{\rm in}) + \varphi_{\rm in} \, .
\end{equation}

\subsubsection{The $\tw$-equation}
\label{eqn:w-solution}

Using equation \eqref{eqn:r-equation} and the subsitution $\tr=\pm\frac{b_3}{4y-\frac{b_2}{3}}$ the $\tw$-equation \eqref{eqn:w-equation} becomes
\begin{equation}
 \dd\tw = \frac{Jb_3^2}{16\left(y-\frac{b_2}{12}\right)^2} \frac{\dd y}{\sqrt{P_3(y)}}\, .
\label{eqn:w-diff}
\end{equation}
The right hand side is an elliptic differential of the third kind. After the substitution $y=\wp(v)$ with $\wp^\prime(v)=\sqrt{4 \wp^3(v)-g_2\wp(v)-g_3}$ and integration of equation \eqref{eqn:w-diff} one gets
\begin{equation}
 \tw(\gamma) - \tw_{\rm in}=\frac{Jb_3^2}{16} \int _{v_{\rm in}}^v \!\frac{1}{\left( \wp(v')-\wp(u)\right)^2 } \dd v' \, ,
\end{equation}
where $v=v(\gamma)=\gamma-\gamma^\prime_{\rm in}$,$v_{\rm in}=v(\gamma_{\rm in})$ and $\wp(u)=\frac{b_2}{12}=-\frac{\tL^2}{12}$. The solution of this integral is given in terms of the elliptic $\wp$-, $\sigma$- and $\zeta$-function \cite{Kagramanova:2010bk, Grunau:2010gd}:
\begin{equation}
\begin{split}
 \tw (\gamma) &= -\frac{Jb_3^2}{16}\frac{\wp''(u)}{(\wp'(u))^3}\left( 2\zeta(u)(v-v_{\rm in})+\ln\frac{\sigma(v-u)}{\sigma(v_{\rm in}-u)} -\ln\frac{\sigma(v+u)}{\sigma(v_{\rm in}+u)}\right) \\
              &-\frac{Jb_3^2}{16}\frac{1}{(\wp'(u))^2}\left(2\wp(u)(v-v_{\rm in}) + 2(\zeta(v)-\zeta(v_{\rm in})) + \frac{\wp'(v)}{\wp(v)-\wp(u)} - \frac{\wp'(v_{\rm in})}{\wp(v_{\rm in})-\wp(u)}\right) + \tw_{\rm in} \, .
\end{split}
\end{equation}

\subsubsection{The $\tlt$-equation}

Using equation \eqref{eqn:r-equation} the $\tlt$-equation \eqref{eqn:t-equation} becomes
\begin{equation}
 \dd\tlt=\frac{E\tr^3}{\tr-1}\frac{\dd\tr}{\sqrt{R(\tr)}}\, .
\end{equation}
Now we substitute $\tr=\pm\frac{b_3}{4y-\frac{b_2}{3}}$ and apply a partial fractions decomposition. Then we get
\begin{equation}
 \dd\tlt = \left( \sum^2_{i=1}\frac{K_i}{y-p_i} + \frac{K_1^2}{(y-p_1)^2}\right) \frac{\dd y}{\sqrt{P_3(y)}},
\label{eqn:t-diff}
\end{equation}
where $p_1=-\frac{\tL^2}{12}$, $p_2=\frac{\tL^2}{6}$, $K_1=\frac{E\tL^2}{4}$ and $K_2=-\frac{E\tL^2}{4}$. The right hand side consists of elliptic differentials of the third kind. After the substitution $y=\wp(v)$ with $\wp^\prime(v)=\sqrt{4 \wp^3(v)-g_2\wp(v)-g_3}$ and integration of equation \eqref{eqn:t-diff} one gets
\begin{equation}
 \tlt(\gamma) - \tlt_{\rm in} =  \int _{v_{\rm in}}^v \!\left( \sum^2_{i=1}\frac{K_i}{\wp(v')-\wp(v_i)} + \frac{K_1^2}{(\wp(v')-\wp(v_1))^2}\right) \, \dd v' \, ,
\end{equation}
where $v=v(\gamma)=\gamma-\gamma^\prime_{\rm in}$,$v_{\rm in}=v(\gamma_{\rm in})$ and $p_i=\wp(v_i)$. The solution of this integral is given in terms of the elliptic $\wp$-, $\sigma$- and $\zeta$-function \cite{Kagramanova:2010bk, Grunau:2010gd}:
\begin{equation}
\begin{split}
  \tlt(\gamma) &= \sum_{i=1}^2 \frac{K_i}{\wp'(v_i)} \left( 2\zeta(v_i)(v-v_{\rm in})+\ln\frac{\sigma(v-v_i)}{\sigma(v_{\rm in}-v_i)} -\ln\frac{\sigma(v+v_i)}{\sigma(v_{\rm in}+v_i)}\right)\\
      & -K_1\frac{\wp''(v_1)}{(\wp'(v_1))^3}\left( 2\zeta(v_1)(v-v_{\rm in})+\ln\frac{\sigma(v-v_1)}{\sigma(v_{\rm in}-v_1)} -\ln\frac{\sigma(v+v_1)}{\sigma(v_{\rm in}+v_1)}\right) \\
      &-\frac{K_1}{(\wp'(v_1))^2}\left(2\wp(v_1)(v-v_{\rm in}) + 2(\zeta(v)-\zeta(v_{\rm in})) + \frac{\wp'(v)}{\wp(v)-\wp(v_1)} - \frac{\wp'(v_{\rm in})}{\wp(v_{\rm in})-\wp(v_1)}\right) + \tlt_{\rm in} \, .
\end{split}
\end{equation}

\subsection{The orbits}

\begin{figure}[ht]
 \centering
 \subfigure[$x$-$y$-$z$-plot]{
   \includegraphics[width=6cm]{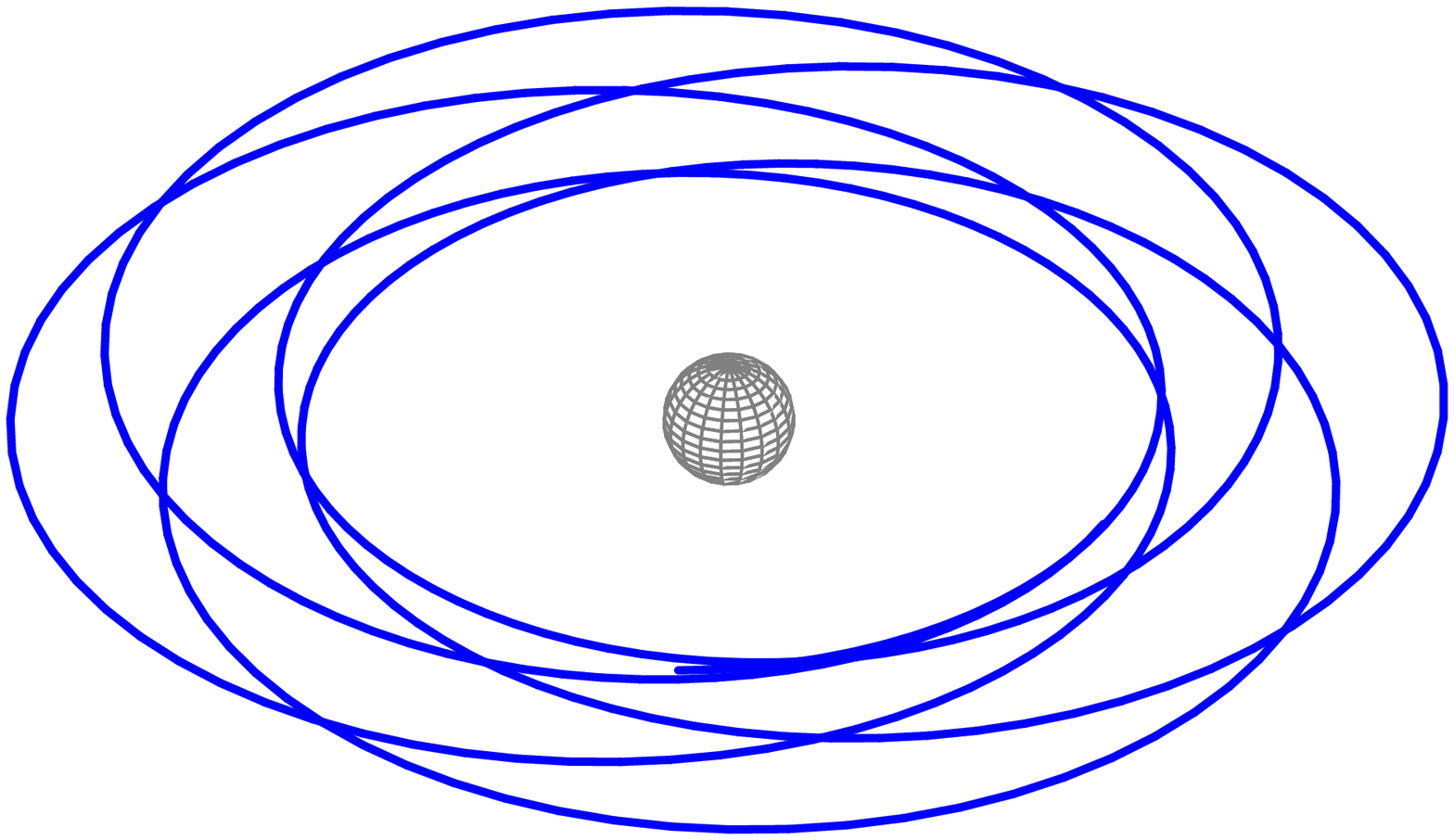}
 }
 \subfigure[$x$-$y$-$w$-plot]{
   \includegraphics[width=6cm]{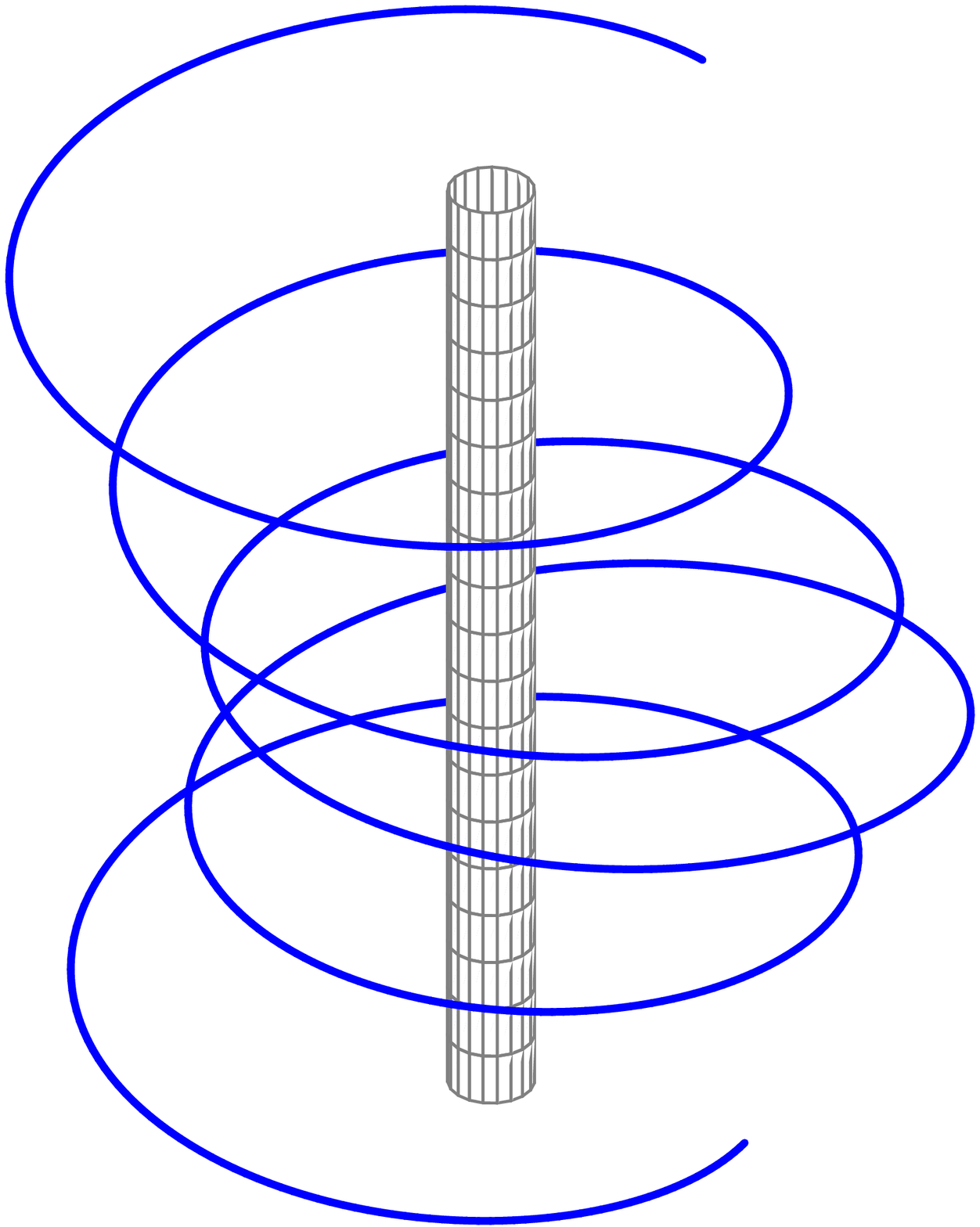}
 }
 \caption{$\delta=0$, $\tL=\sqrt{5}$, $J=1$ and $E=0.973$:\newline
          Bound orbit for light in the black string spacetime. This orbit is not possible in the ordinary four-dimensional Schwarzschild  spacetime. The sphere or cylinder is the horizon.}
 \label{pic:bo-light}
\end{figure}

\begin{figure}[ht]
 \centering
 \subfigure[$x$-$y$-$z$-plot]{
   \includegraphics[width=6cm]{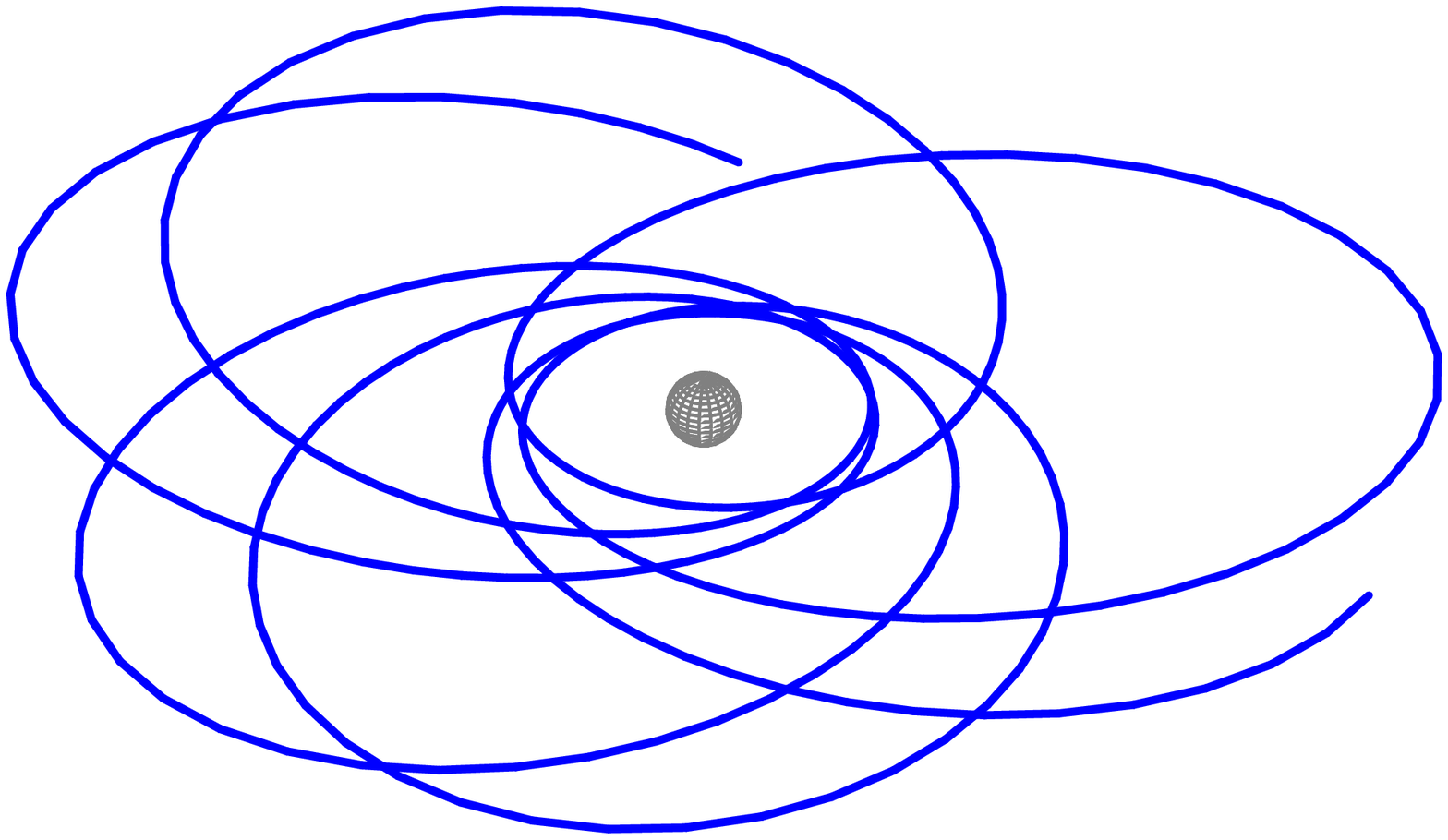}
 }
 \subfigure[$x$-$y$-$w$-plot]{
   \includegraphics[width=6cm]{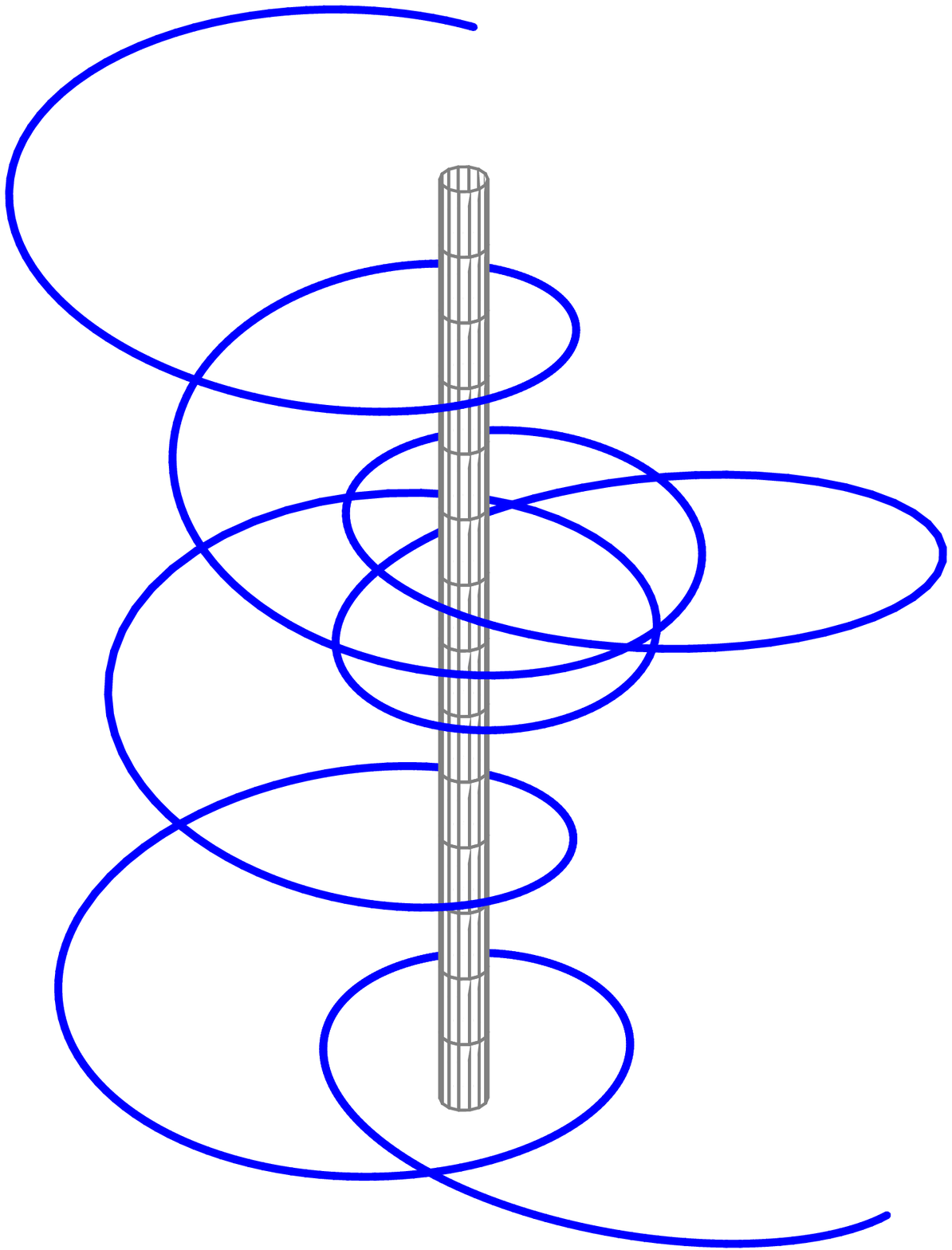}
 }
 \caption{$\delta=1$, $\tL=\sqrt{5}$, $J=0.2$ and $E=1$:\newline
          Bound orbit for particles in the black string spacetime. The sphere or cylinder is the horizon.}
 \label{pic:bo-particles}
\end{figure}

\begin{figure}[ht]
 \centering
 \subfigure[$x$-$y$-$z$-plot]{
   \includegraphics[width=6cm]{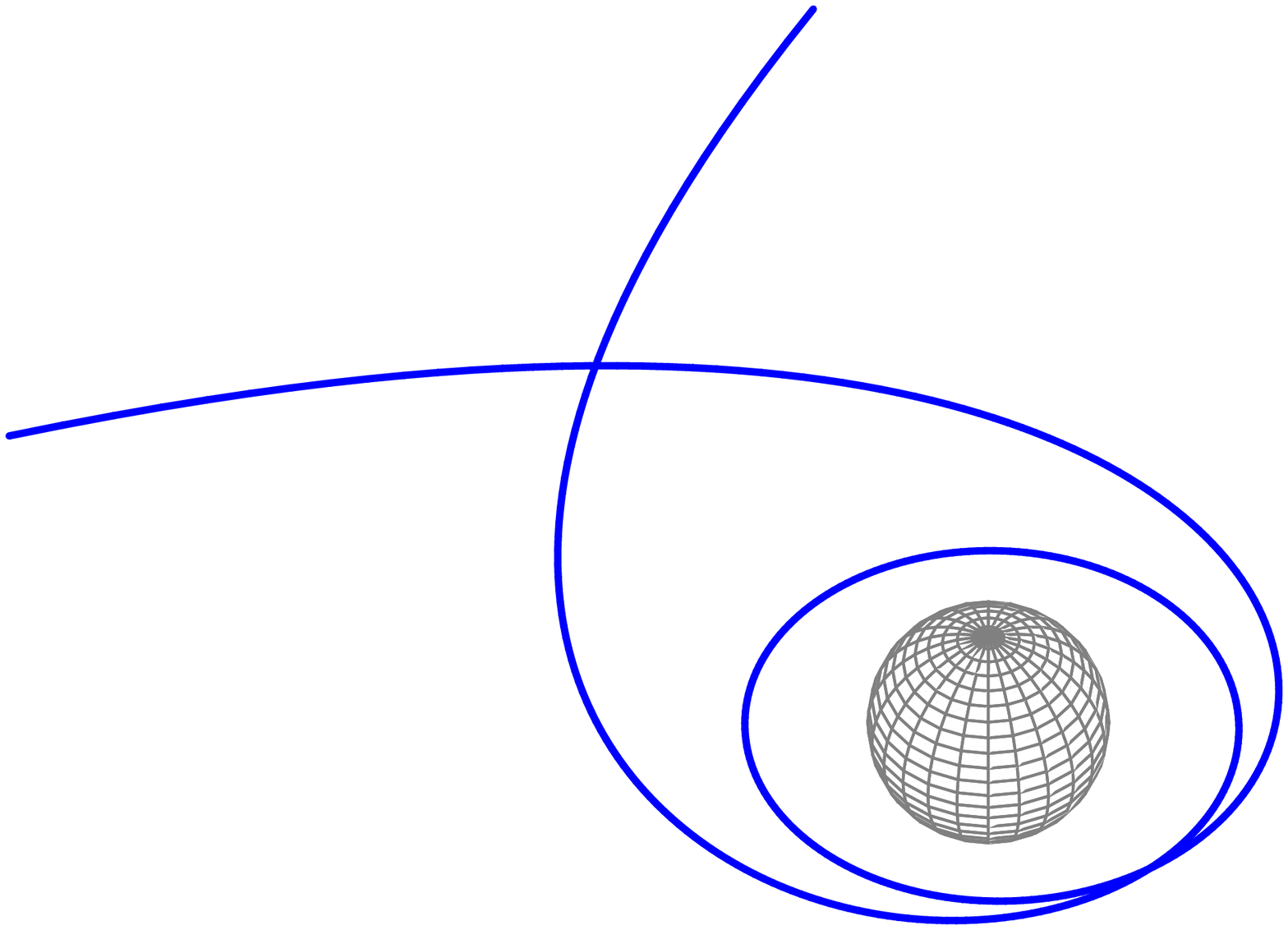}
 }\qquad\qquad
 \subfigure[$x$-$y$-$w$-plot]{
   \includegraphics[width=6cm]{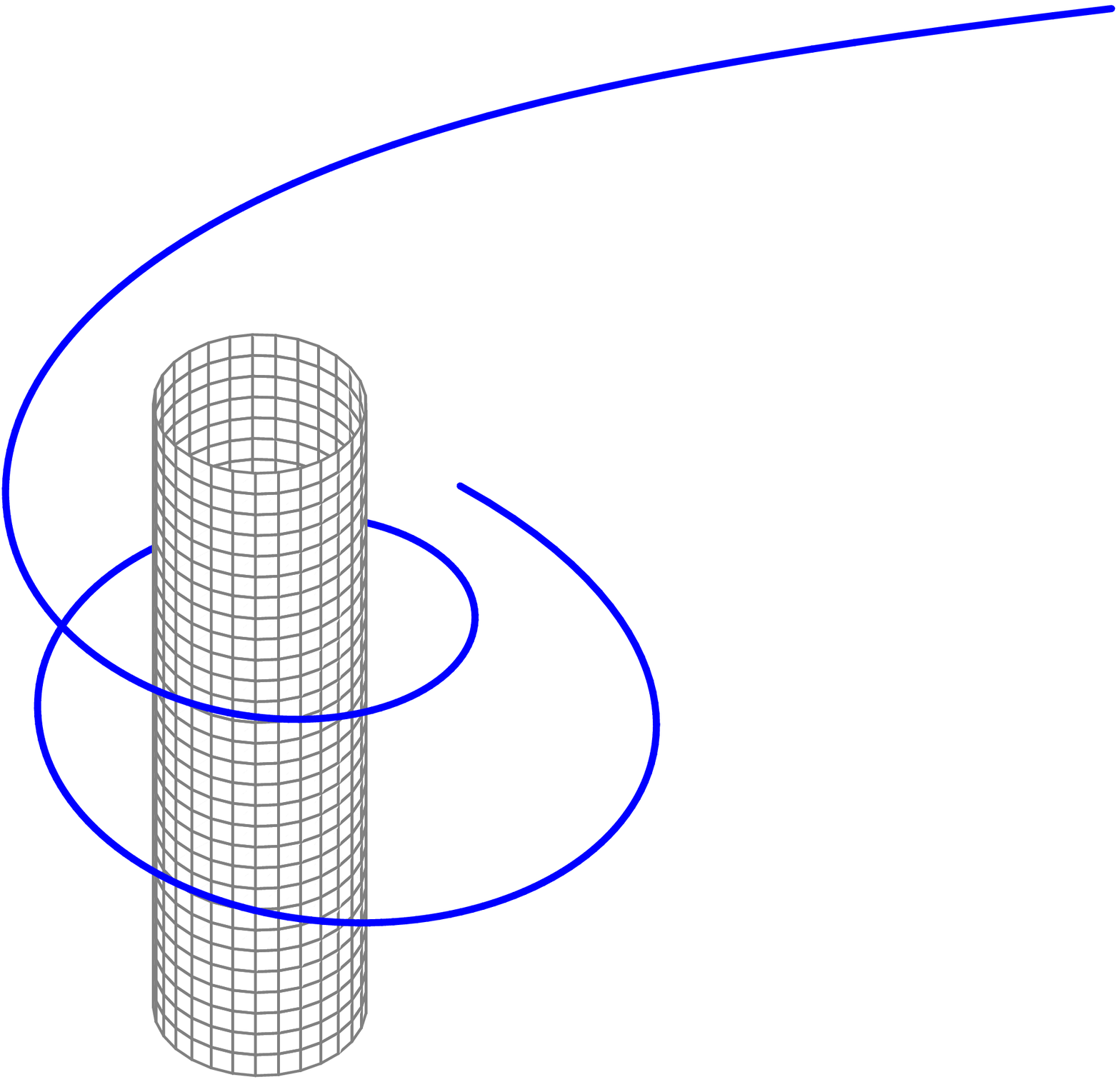}
 }
 \caption{$\delta=0$, $\tL=\sqrt{5}$, $J=1.1$ and $E=1.109$:\newline
          Escape orbit for light in the black string spacetime. The sphere or cylinder is the horizon.}
 \label{pic:eo-light}
\end{figure}

\begin{figure}[ht]
 \centering
 \subfigure[$x$-$y$-$z$-plot]{
   \includegraphics[width=6cm]{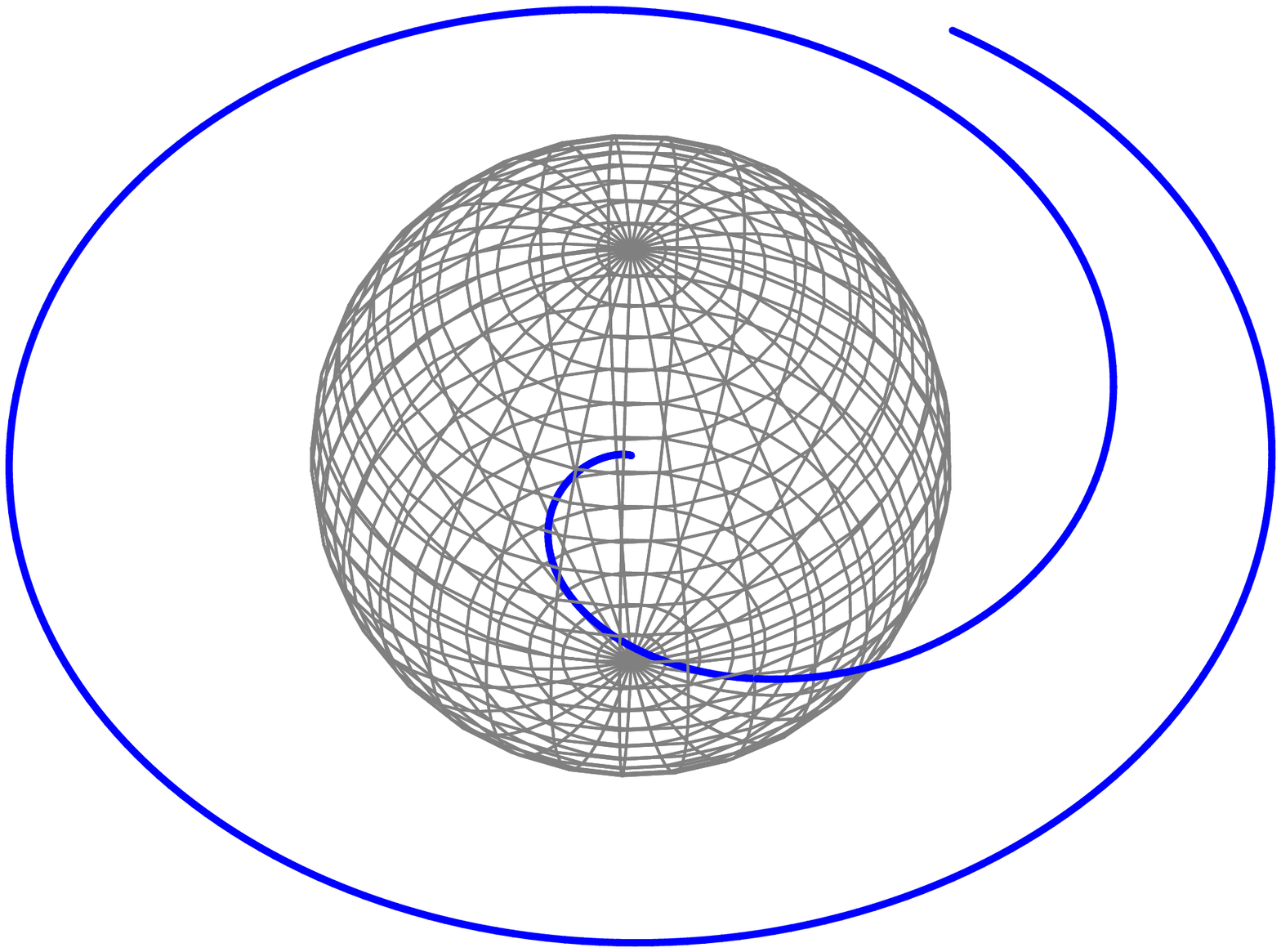}
 }\qquad\qquad
 \subfigure[$x$-$y$-$w$-plot]{
   \includegraphics[width=6cm]{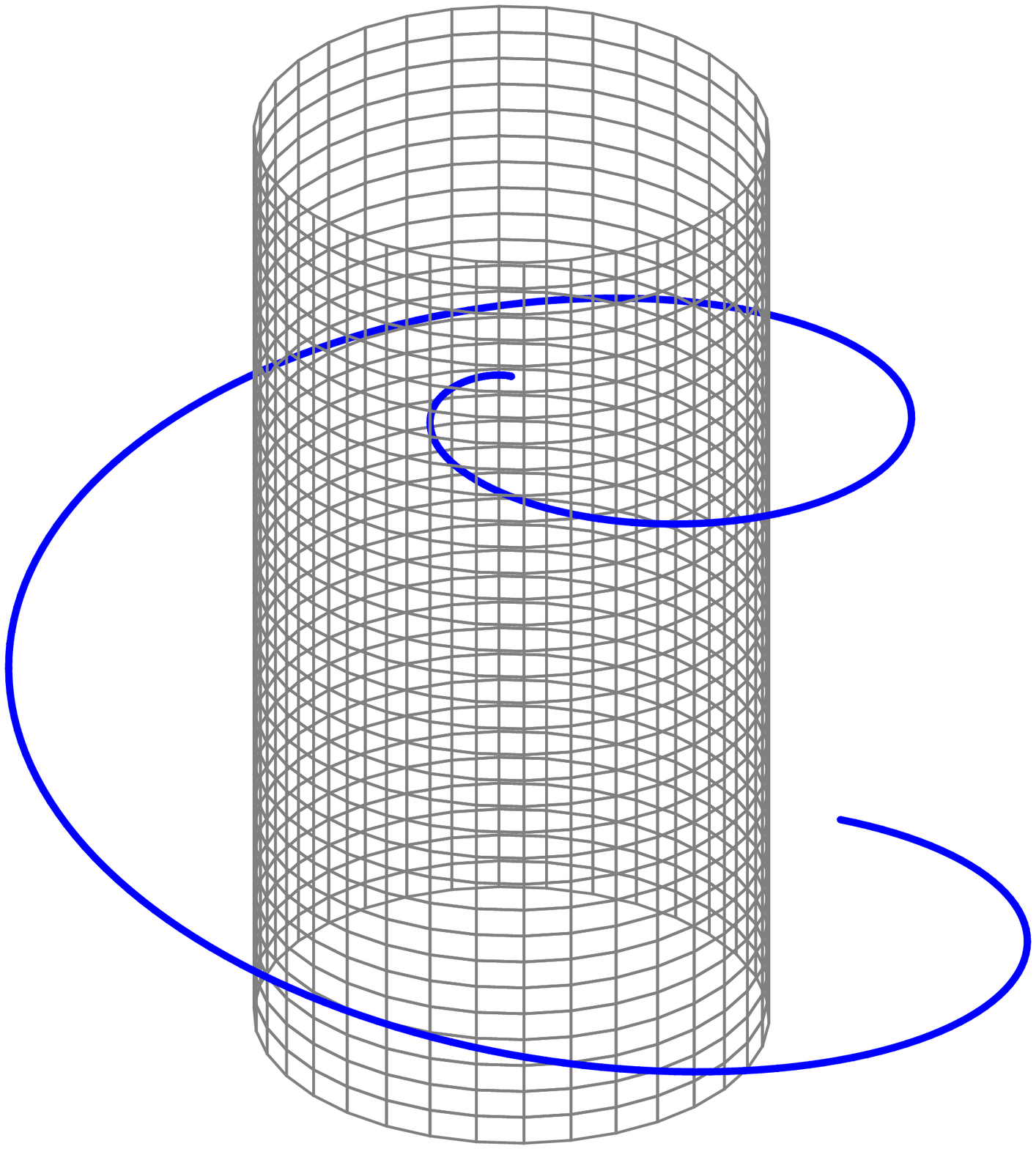}
 }
 \caption{$\delta=1$, $\tL=2$, $J=0.1$ and $E=1.0025$:\newline
          Terminating orbit for particles in the black string spacetime. The sphere or cylinder is the horizon.}
 \label{pic:to-particles}
\end{figure}

With these analytical results we have found the complete set of orbits for light and test particles in the black string spacetime. Depending on the parameters $\delta$, $\tL$, $J$ and $E$, BOs, EOs, and TOs are possible. BOs can be seen in figure \ref{pic:bo-light} and \ref{pic:bo-particles}. An EO is depicted in figure \ref{pic:eo-light}, and figure \ref{pic:to-particles} shows a TO. In Cartesian $x$-$y$-$z$ coordinates the orbits lie in a plane. However, this is not the case if the orbit is plotted in $x$-$y$-$w$ coordinates.

In the black string spacetime BOs for light are possible (see figure \ref{pic:bo-light}). Such orbits do not exist in the ordinary four-dimensional Schwarzschild spacetime.

\section{Rotating black string spacetime}

In this section we will discuss the geodesics in the rotating black string spacetime and present analytical solutions of the equations of motion. The rotating black string metric is derived from the Kerr metric by adding an extra compact dimension. A detailed analysis of the geodesics in the original Kerr spacetime without a compact dimension can be found in e.g. \cite{ONeill:1995}.

\subsection{The geodesic equations}

The Kerr metric is the rotating axially symmetric solution of the Einstein equations in four dimensions. If an extra compact spatial dimension $w$ is added, the metric takes the form
\begin{equation}
 \dd s^2 = -\frac{\Dr}{\rho^2}(\dd t-a\sin^2\vartheta\dd\varphi)^2 + \frac{\sin^2\vartheta}{\rho^2}((r^2+a^2)\dd\varphi-a\dd t)^2 + \frac{\rho^2}{\Dr}\dd r^2 + \rho^2\dd\vartheta^2 + \dd w^2
\end{equation}
where $\Dr= r^2-2Mr+a^2$ and $\rho^2=r^2+a^2\cos^2\vartheta$. This solution describes a rotating uniform black string. $M$ is proportional to the mass of the black string and $a$ is proportional to the angular momentum. The gravitational constant and the velocity of light are set to 1. The (ring) singularity is located at $\rho^2=0$, i.e. at $r=0$ and $\vartheta=\frac{\pi}{2}$. Hence geodesics with  $r=0$ and $\vartheta\neq\frac{\pi}{2}$ do not end in the singularity. So in the rotating black string spacetime negative $r$-values are allowed as in the Kerr spacetime \cite{ONeill:1995}. There are two horizons defined by $\Dr=0$ and for $a^2<\frac{1}{4}$ they are given by
\begin{equation}
 r_\pm = M \pm \sqrt{ M^2 - a^2} \, .
\end{equation}
Note that the metric is given in Boyer-Lindquist coordinates $r$, $\vartheta$, $\varphi$. Cartesian coordinates can be obtained by the transformation
\begin{equation}
 \begin{split}
  x&=\sqrt{(r^2+a^2)}\sin\vartheta\cos\varphi \\
  y&=\sqrt{(r^2+a^2)}\sin\vartheta\sin\varphi \\
  z&=r\cos\vartheta \, .
 \end{split}
\end{equation}

The Hamilton-Jacobi equation
\begin{equation}
 \frac{\partial S}{\partial \tau} + \half g^{\mu\nu} \frac{\partial S}{\partial x^\mu}\frac{\partial S}{\partial x^\nu}=0
\label{eqn:ham-jac2}
\end{equation}
has a solution of the form
\begin{equation}
 S=\half\delta \tau -Et+L\varphi + Jw + S_r(r) + S_\vartheta (\vartheta) \, .
\end{equation}
Here $\tau$ is an affine parameter along the geodesic. The parameter $\delta$ is equal to $1$ for particles and equal to $0$ for light. $E$, $L$ and $J$ are the conserved momenta; $E$ is the energy, $L$ denotes the angular momentum and $J$ is a new constant of motion according to the compact dimension $w$.

For convenience, we introduce dimensionless quantities ($r_{\rm S}=2M$)
\begin{equation}
\tr=\frac{r}{r_{\rm S}} \ , \,\, \tw=\frac{w}{r_{\rm S}} \ , \,\, \tlt=\frac{t}{r_{\rm S}} \ , \,\, \tilde{\tau}=\frac{\tau}{r_{\rm S}} \ , \,\, \tL=\frac{L}{r_{\rm S}} \ , \,\, \ta=\frac{a}{r_{\rm S}} \ , \,\, \tK=\frac{K}{r_{\rm S}}  \ .
\end{equation}
$K$ is the famous Carter constant resulting from the separation of the Hamilton-Jacobi equation, see \cite{Carter:1968rr}. In these dimensionless quantities the horizons are located at $\tr_\pm = \half \pm \sqrt{\frac{1}{4}-\ta^2}$.

The Hamilton-Jacobi equation \eqref{eqn:ham-jac} separates and yields a differential equation for each coordinate
\begin{eqnarray}
 \left(\frac{\dd \tr}{\dd \gamma}\right) ^2 &=& R \label{eqn:kerr-r-equation}\\
 \left(\frac{\dd \vartheta}{\dd \gamma}\right) ^2 &=& \Theta \label{eqn:kerr-theta-equation}\\
 \frac{\dd \varphi}{\dd \gamma} &=& \frac{\ta}{\tDr}[(\tr^2+\ta^2)E-\ta\tL] - \frac{1}{\sin^2\vartheta}(\ta E \sin^2\vartheta -\tL) \label{eqn:kerr-phi-equation}\\
 \frac{\dd \tw}{\dd \gamma} &=& J\trho ^2 \label{eqn:kerr-w-equation} \\
 \frac{\dd \tlt}{\dd \gamma} &=& \frac{\tr^2+\ta^2}{\tDr}[(\tr^2+\ta^2)E-\ta\tL] - \ta(\ta E \sin^2\vartheta -\tL)  \label{eqn:kerr-t-equation}
\end{eqnarray}
with the polynomial $R$ and the function $\Theta$:
\begin{equation}
 \begin{split}
  R &=[(\tr^2+\ta^2)E-\ta\tL]^2 - \tDr(\tK+(\delta+J^2)\tr^2) \\
  \Theta &= \tK - (\delta +J^2)\ta^2\cos^2\vartheta - \frac{1}{\sin^2\vartheta}(\ta E \sin^2\vartheta -\tL)^2  \, .
 \end{split}
\end{equation}
We also used the Mino time $\gamma$ as $\trho^2\dd\gamma = \dd\tilde{\tau}$ \cite{Mino:2003yg}. The new constant of motion $J$ due to the compact dimension is present in the $\tr$-equation \eqref{eqn:kerr-r-equation}, the $\vartheta$-equation \eqref{eqn:kerr-theta-equation} and of course in the $\tw$-equation \eqref{eqn:kerr-w-equation}. The other equations are the same as in the original Kerr spacetime without the compact dimension. To obtain the geodesic equations of the original Kerr spacetime, one can choose $J=0$.

\subsection{Classification of the geodesics}

The function $\Theta$ and the polynomial $R$ define the properties of the orbits depending on the parameters of the metric and the test particle. In this section we will analyze the function $\Theta$ and the polynomial $R$ to determine the possible types of orbits. A similar analysis of the Kerr spacetime can be found in \cite{ONeill:1995}, for the Kerr-(anti-) de Sitter spacetime see \cite{Hackmann:2010zz}.\\

To obtain real values of $\vartheta$ and $\tr$ the conditions $\Theta\geq0$ and $R\geq0$ have to be fullfilled, otherwise no motion is possible. From $\Theta\geq0$ we can immediately conclude that $\tK\geq0$. If $\tK=0$, the motion takes place in the equatorial plane \cite{ONeill:1995}.

Furthermore we see that a geodesic hits the singularity at $\trho^2=0$ if $\tK =(\ta E-\tL)^2$:
A solution of $\trho^2=\tr^2+\ta^2\cos^2\vartheta=0$ is $\tr=0$ and simultaneously $\vartheta=\frac{\pi}{2}$. Since
\begin{equation}
 \begin{split}
  \Theta\left(\frac{\pi}{2}\right)&=\tK-(\ta E-\tL)^2\geq0 \quad \text{and} \\
  R(0)&=-\ta^2[\tK-(\ta E-\tL)^2] \geq0 \, ,
 \end{split}
\label{eqn:kerr-R-Theta-0}
\end{equation}
it follows that $\tK=(\ta E-\tL)^2$. Additionally an orbit lies in the equatorial plane ($\vartheta=\frac{\pi}{2}$) if $\tK=(\ta E-\tL)^2$ (compare\cite{Hackmann:2010zz,ONeill:1995}). From \eqref{eqn:kerr-R-Theta-0} we also see the following:
\begin{itemize}
 \item $\tK>(\ta E-\tL)^2$: The geodesics cross $\vartheta=\frac{\pi}{2}$ but do not cross $\tr=0$.
 \item $\tK<(\ta E-\tL)^2$: The geodesics do not cross $\vartheta=\frac{\pi}{2}$ but it is possible to cross $\tr=0$.
\end{itemize}
Here the rotating black string shows the same properties as the Kerr black hole (see \cite{ONeill:1995}).

\subsubsection{The $\vartheta$-motion}

As in \cite{Hackmann:2010zz} we substitute $\nu=\cos^2\vartheta$ (with $\nu\in[0,1]$) in the function $\Theta(\vartheta)$:
\begin{equation}
 \Theta(\nu)=\tK-(\delta+J^2)\ta^2\nu-\ta^2 E^2(1-\nu)+2\ta E\tL -\frac{\tL^2}{(1-\nu)} \, .
\end{equation}
Now we want to determine the number of real zeros of $\Theta(\nu)$ in $[0,1]$. The number of zeros only changes if a zero crosses 0 or 1, or if a double zero occurs. $\nu=0$ is a zero of $\Theta$ if
\begin{equation}
 \Theta(\nu=0)=\tK-(\ta E-\tL)^2=0
\end{equation}
and therefore
\begin{equation}
 \tL=\ta E\pm\sqrt{\tK} \, .
 \label{eqn:kerr-theta-border1}
\end{equation}
Since $\nu=1$ is a pole of $\Theta(\nu)$ for $\tL\neq 0$, it is only possible that $\nu=1$ is a zero of $\Theta(\nu)$ if $\tL=0$.
\begin{equation}
 \Theta(\nu=1, \tL=0)= \tK-(\delta+J^2)\ta^2=0
\end{equation}
yields that $\tK=(\delta+J^2)\ta^2$. Note that due to the new constant of motion $J$ (the momentum in the $\tw$-direction) the value of $\tK$ is different here in comparison to the Kerr spacetime.

So $\tL=\ta E\pm\sqrt{\tK}$ and $\tL=0 \, \wedge \, \tK=(\delta+J^2)\ta^2$ are border cases of the $\vartheta$-motion.\\

Next we consider the polynomial
\begin{equation}
 \Theta'(\nu)= \ta^2(\delta+J^2-E^2)\nu^2 + [2\ta E(\ta E-\tL)-(\delta+J^2)\ta^2-\tK]\nu + \tK-(\ta E-\tL)^2
\end{equation}
to remove the pole of $\Theta(\nu)$ at $\nu=1$. $\Theta$ and $\Theta'$ are related by $\Theta(\nu)=\frac{1}{1-\nu}\Theta'(\nu)$. Double zeros of $\Theta(\nu)$ in $[0,1)$ and hence of $\Theta'(\nu)$ occur if
\begin{equation}
 \Theta'(\nu)=0 \quad \text{and} \quad \frac{\dd \Theta' (\nu)}{\dd\nu}=0 \, .
\end{equation}
From these conditions we obtain
\begin{equation}
 \tL=\frac{E\pm (\ta^2(\delta+J^2)-\tK)\sqrt{E^2-(\delta+J^2)}}{2\ta (\delta+J^2)} \, .
  \label{eqn:kerr-theta-border2}
\end{equation}
Therefore double zeros are only possible if $E^2\geq\delta+J^2$.\\

\begin{figure}[h]
 \centering
 \includegraphics[width=6cm]{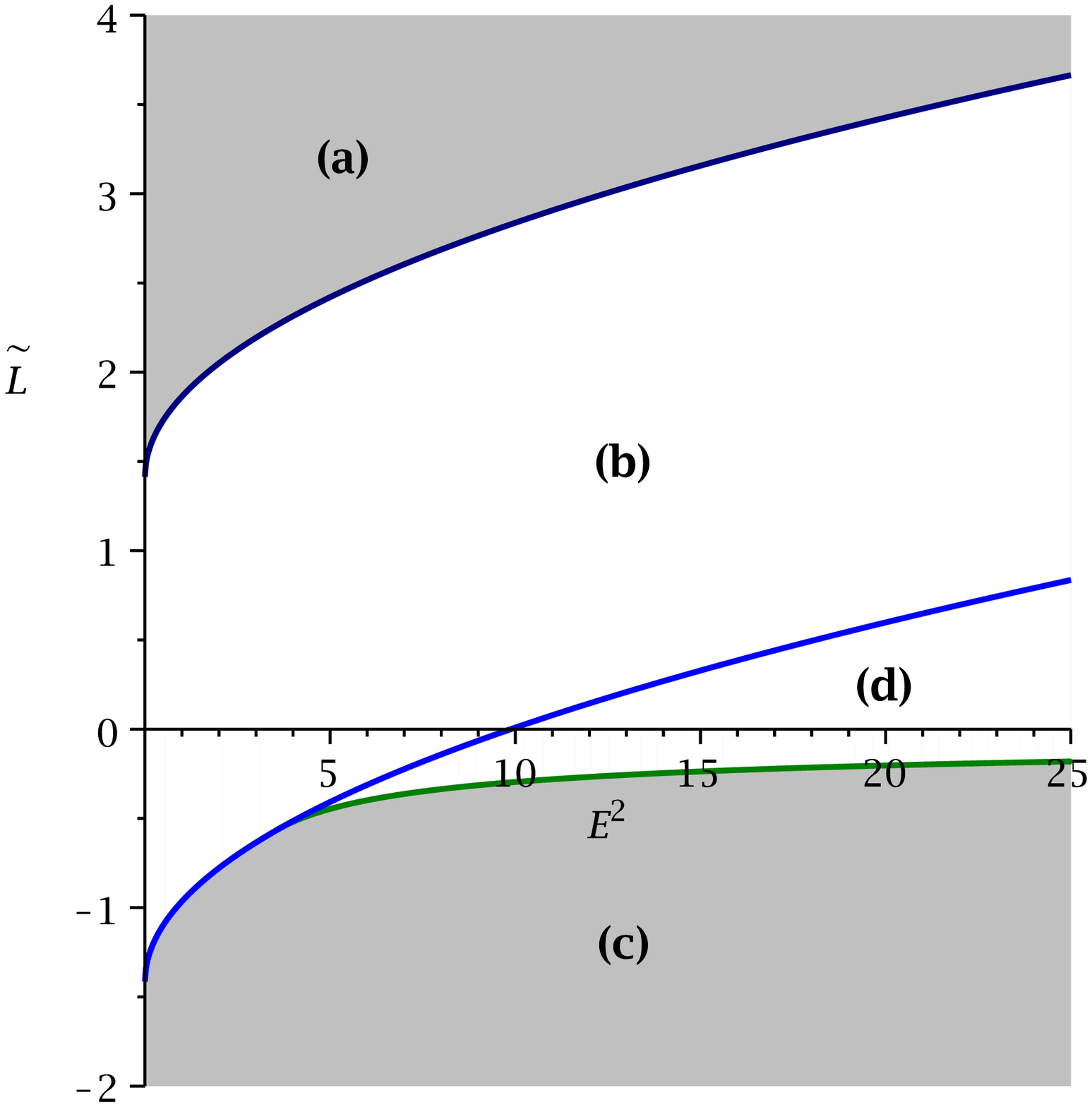}
 \caption{$\delta=1$, $\ta=0.45$, $J=1$, $\tK=2$: \newline
          Parametric $\tL$-$E^2$-diagram of the $\vartheta$-motion. In the grey regions (a) and (d) we have $\Theta(\nu)<0$ for all $\nu\in[0,1]$ and hence no motion is possible. In region (b) $\Theta(\nu)$ has one real zero in $[0,1]$ and in region (d)  $\Theta(\nu)$ has two real zeros in $[0,1]$.}
 \label{pic:kerr-theta-parametric}
\end{figure}

With the help of equation \eqref{eqn:kerr-theta-border1} and \eqref{eqn:kerr-theta-border2} parametric $\tL$-$E^2$-diagrams can be drawn. Figure \ref{pic:kerr-theta-parametric} shows a typical example of a $\tL$-$E^2$-diagram in the rotating black string spacetime, which resembles the parametric diagrams in the Kerr or Kerr-(anti-) de Sitter spacetime \cite{Hackmann:2010zz}. Here we recognize four regions. In region (a) and (d) we have  $\Theta(\nu)<0$ for all $\nu\in[0,1]$ and therefore no geodesic motion is possible. In region (b)  $\Theta(\nu)$ has one real zero in $[0,1]$, here $\tK>(\ta E-\tL)^2$ and the orbit crosses $\vartheta=\frac{\pi}{2}$. In region (d)  $\Theta(\nu)$ has two real zeros in $[0,1]$, here $\tK<(\ta E-\tL)^2$ and $\vartheta=\frac{\pi}{2}$ is not crossed. However, crossing $\tr=0$ is possible in region (d) but not in (b).\\

In the special case $\tL=0$ there is a single zero at $\nu_0=\frac{\ta^2E^2-\tK}{\ta^2(E^2-(\delta+J^2))}$ (not necessarily in $[0,1]$). If additionally $\tK=(\delta+J^2)\ta^2$ then this zero is at $\nu=1$ and if $\tK=\ta^2E^2$ this zero is at $\nu=0$. In the case $\tK>(\delta+J^2)\ta^2$ there is a zero $\nu_0$ in $[0,1]$ if $\tK<\ta^2E^2$ and $\Theta(\nu)>0$ for all $\nu\in[0,1]$ if $\tK>\ta^2E^2$. In the case $\tK<(\delta+J^2)\ta^2$ there is a zero $\nu_0$ in $[0,1]$ if $\tK>\ta^2E^2$ but $\Theta(\nu)<0$ for all $\nu\in[0,1]$ if $\tK<\ta^2E^2$.

\subsubsection{The $\tr$-motion}

The zeros of the polynomial $R$ are the turning points of orbits of light and test particles and therefore $R$ determines the possible types of orbits. In contrast to the non-rotating black string, here we can consider negatives $\tr$-value too, since there is no singularity for $r=0$ and $\vartheta\neq\frac{\pi}{2}$. Orbits that cross $\tr=0$ are called crossover orbits \cite{Hackmann:2010zz}. Before discussing the $\tr$-motion in detail, we introduce a list of all possible orbits:

\begin{enumerate}
 \item \textit{Transit orbit} (TrO) with range $\tr \in (-\infty, \infty)$. This orbit is a crossover orbit.
 \item \textit{Escape orbit} (EO) with range $\tr \in [r_1, \infty)$ with $r_1>\tr_+$, or with range $\tr \in (-\infty, r_1]$ with  $r_1<0$.
 \item \textit{Two-world escape orbit} (TEO) with range $[r_1, \infty)$ where $0<r_1 < r_-$. The TEOs cross both horizons twice and emerge into another universe.
 \item \textit{Crossover two-world escape orbit} (CTEO) with range $[r_1, \infty)$ where $r_1 < 0$. The CTEOs cross both horizons twice and emerge into another universe. $\tr=0$ is crossed once.
 \item \textit{Bound orbit} (BO) with range $\tr \in [r_1, r_2]$ with $0<r_1 < r_2$ and
  \begin{enumerate}
   \item either $r_1, r_2  > r_+$ or 
   \item $r_1, r_2 < r_-$.
  \end{enumerate}
 \item \textit{Many-world bound orbit} (MBO) with range $\tr \in [r_1, r_2]$ where $0<r_1 \leq r_-$ and $r_2 \geq r_+$. The MBOs cross both horizons several times. Each time both horizons are traversed twice the test particles emerge into another universe.
 \item \textit{Terminating orbit} (TO) with ranges $\tr \in [0, \infty)$ or $\tr \in [0, r_1]$ with
  \begin{enumerate}
   \item either $r_1\geq\tr_+$ or 
   \item $0<r_1<\tr_-$.
  \end{enumerate}
  The TOs end in the singularity at $\tr=0$ and $\vartheta=\frac{\pi}{2}$. $\tK=(\ta E-\tL)^2$ is required.
\end{enumerate}

\begin{figure}[h]
 \centering
 \includegraphics[width=6cm]{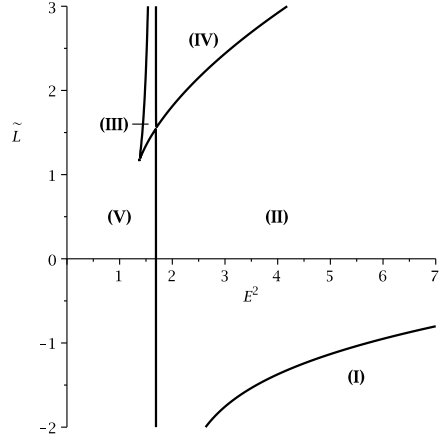}
 \caption{$\delta=0$, $\ta=0.45$, $J=1.3$, $\tK=2$: \newline
          Parametric $\tL$-$E^2$-diagram of the $\tr$-motion. In region (I) $R$ has no real zeros. In the regions (II) and (V) $R$ has two zeros. Four zeros are possible in the regions (III) and (IV). Here we see that in the rotating black string spacetime all five regions are present for $\delta=0$, whereas in the Kerr spacetime region (III) and (V) do not exist if $\delta=0$.}
 \label{pic:kerr-r-parametric}
\end{figure}

\begin{figure}[h]
 \centering
 \subfigure[$\delta=1$, $\ta=0.45$, $J=1$, $\tK=2$]{
   \includegraphics[width=5cm]{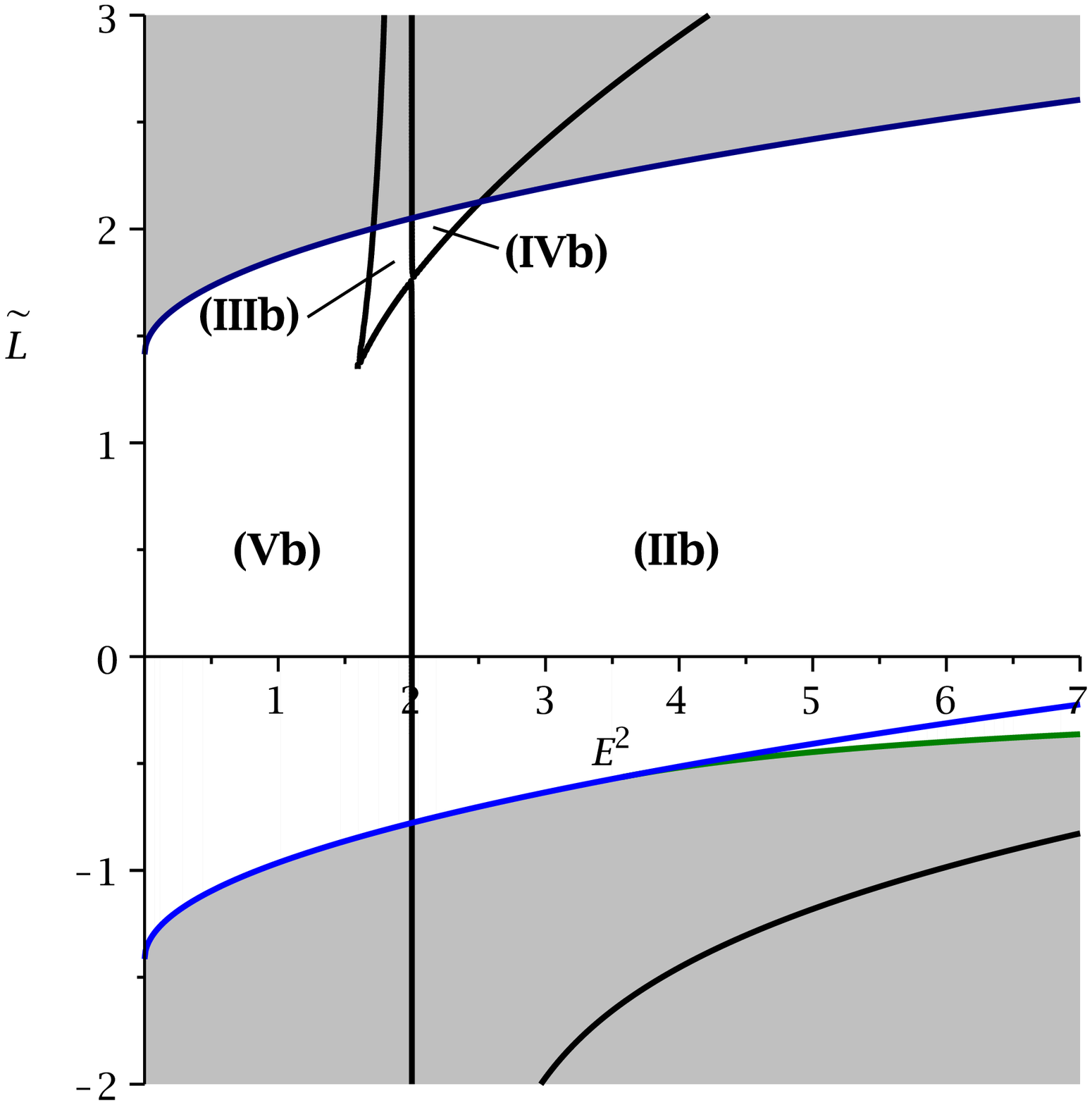}
 }
  \subfigure[$\delta=1$, $\ta=0.45$, $J=1$, $\tK=2$:\newline
  Different view of figure (a), here region (Id) is visible.]{
   \includegraphics[width=5cm]{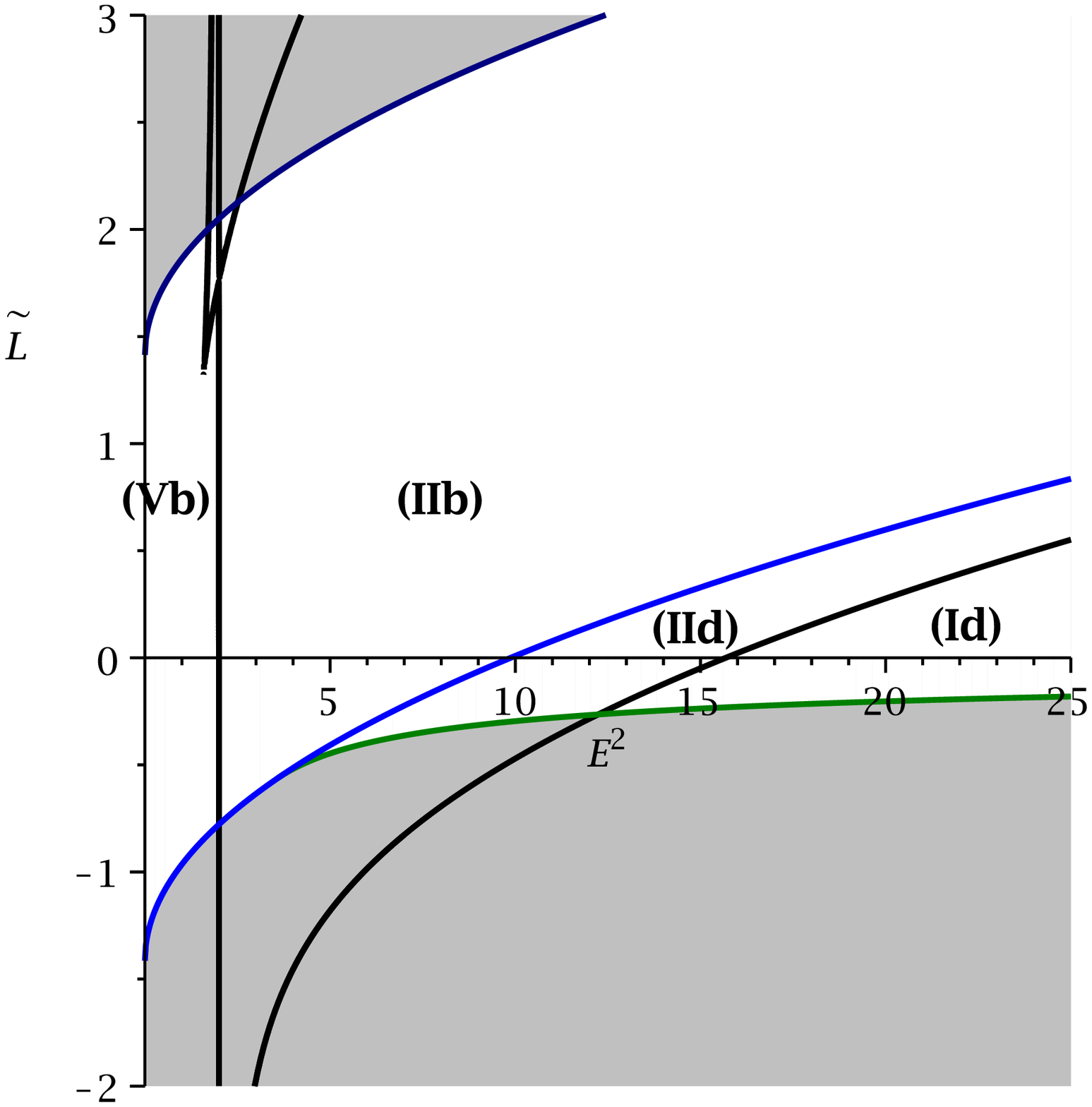}
 }
  \subfigure[$\delta=1$, $\ta=0.45$, $J=0.1$, $\tK=0.2$:\newline 
 For small $\tK$ a second part of region (IVb) appears.]{
   \includegraphics[width=5cm]{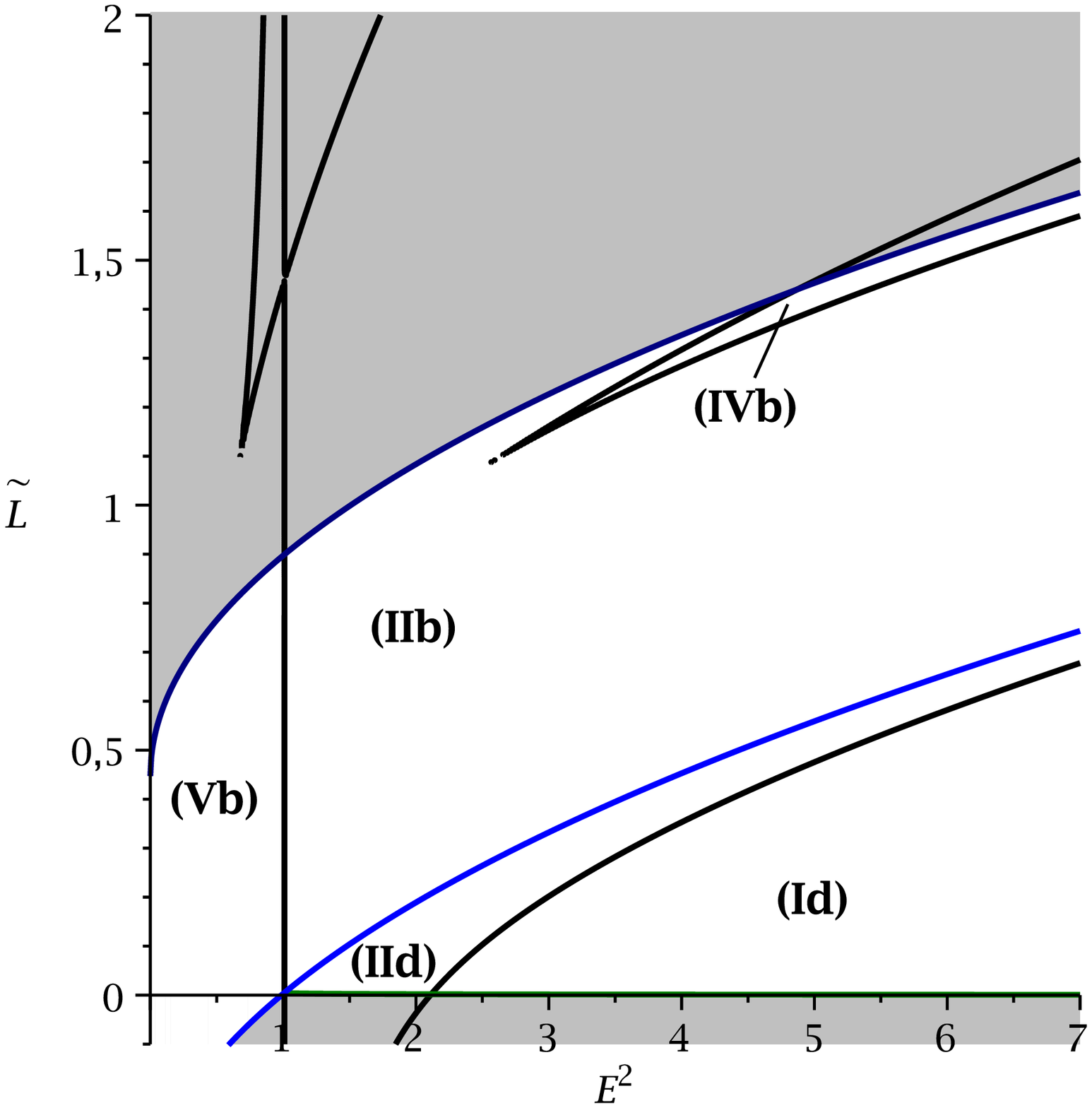}
 }
 \caption{Combined parametric $\tL$-$E^2$-diagram of the $\vartheta$-motion and the $\tr$-motion. In region (I) $R$ has no real zeros. In the regions (II) and (V) $R$ has two zeros. Four zeros are possible in the regions (III) and (IV). In the grey area the $\vartheta$-equation does not allow geodesic motion. In region (b) the geodesics cross $\vartheta=\frac{\pi}{2}$, but $\tr=0$ cannot be crossed. However, in region (d) $\tr=0$ can be crossed but $\vartheta=\frac{\pi}{2}$ is not crossed.}
 \label{pic:kerr-r-theta-parametric}
\end{figure}

The possible orbit types depend on the number of real zeros of the polynomial $R$. The number of zeros can only change if double zeros occur, i.e. if
\begin{equation}
 R(\tr)=0 \quad \text{and} \quad \frac{\dd R(\tr)}{\dd\tr}=0 \, .
 \label{eqn:kerr-r-doublezero}
\end{equation}
With the help of equation \eqref{eqn:kerr-r-doublezero} parametric $\tL$-$E^2$-diagrams can be drawn, where five regions with a different number of real zeros of $R$ appear (see figure \ref{pic:kerr-r-parametric}). Also the $\vartheta$-equation has to be taken into account. Both parametric $\tL$-$E^2$-diagrams for the  $\vartheta$-motion and the $\tr$-motion have to be combined in order to detect all possible behaviour of light and test particles in the rotating black string spacetime (see figure \ref{pic:kerr-r-theta-parametric}).

Additionally one may define an effective potential consisting of the two parts $V^+$ and $V^-$ by the equation
\begin{equation}
 R(\tr)=(\tr^2+\ta^2)^2(E-V^+)(E-V^-) \, ,
\end{equation}
thus
\begin{equation}
 V^\pm = \frac{\ta\tL\pm\sqrt{\Dr [\tK+\tr^2(\delta+J^2)]}}{\tr^2+\ta^2} \, .
\end{equation}
The area between $V^+$ and $V^-$ is a forbidden zone, since there $R(\tr)<0$. $V^+$ and $V^-$ meet at the horizons $\tr_\pm$, where $V^\pm(\tr_\pm)=\frac{\ta \tL}{\tr_\pm}$. In the limit $\tr\rightarrow\infty$ and $\tr\rightarrow -\infty$ the effective potential $V^\pm$ converges to $\pm\sqrt{\delta+J^2}$. If the sign of $\tL$ changes, the effective potential is mirrored at the $\tr$-axis.\\

Taking all informations into acount, we can now determine the possible orbits in the different regions of the parametric diagrams (below we always assume that $r_i<r_{i+1}$):
\begin{enumerate}
 \item Region (I): No real zeros and $R(\tr)>0$ for all $\tr$. Here only transit orbits are possible which cross $\tr=0$. Since region (I) intersects with region (d), $\vartheta=\frac{\pi}{2}$ is not crossed.
 \item Region (II): $R(\tr)$ has two real zeros $r_1$, $r_2$ and $R(\tr)\geq0$ for $\tr\in(-\infty,r_1]$ and $\tr\in[r_2,\infty)$. Region (II) intersects with region (b) and region (d) from the $\vartheta$-motion. In part (IIb) there is a negative and a positive zero, so that escape orbits are possible for $\tr<0$ and two-world escape orbits are possible for $\tr\geq0$. Here the orbits cross $\vartheta=\frac{\pi}{2}$ plane. The turning point of the TEO can coincide with the inner horizon $\tr_-$. In the special case $\tK=(\ta E-\tL)^2$ the former positive zero is now at $\tr=0$ so that the two-world escape orbit turns into a terminating orbit and the motion takes place in the equatorial plane.

 In part (IId) $R$ has two negative zeros, so there is an escape orbit for $\tr<0$ and a crossover two-world escape orbit. This time $\vartheta=\frac{\pi}{2}$ is not crossed.
 \item Region (III): $R$ has four positive zeros $r_1$, $r_2$, $r_3$, $r_4$ and $R(\tr)\geq0$ for $\tr\in [r_1, r_2]$ and $\tr\in [r_3, r_4]$. Possible orbits are bound orbits with $\tr>\tr_+$ and many-world bound orbits. One or both (for $\tL=0$) of the turnings points of the many-world orbit can coincide with the horizons. Since region (III) intersects with region (b) the orbits cross $\vartheta=\frac{\pi}{2}$.
 \item Region (IV): $R$ has one negative zero $r_1$ and three positive zero $r_2$, $r_3$, $r_4$. $R(\tr)\geq0$ for $\tr\in (-\infty, r_1]$, $\tr\in [r_2, r_3]$ and $\tr\in [r_4, \infty)$. Escape orbits with either $\tr<0$ or $\tr>\tr_+$ and many-world bound orbits are possible. If $\tK$ is small then also bound orbits with $0<\tr<\tr_-$ are possible instead of the many-world bound orbits. These orbits are hidden behind the inner horizon.
In the special case $\tK=(\ta E-\tL)^2$ one of the turning points of the (many-world) bound orbit is now at $\tr=0$ so that the orbit turns into a terminating orbit and the motion takes place in the equatorial plane.
 \item Region (V): $R$ has two positive zeros $r_1$, $r_2$ and $R(\tr)\geq0$ for $\tr\in [r_1, r_2]$. Only many-world bound orbits are possible. One or both (for $\tL=0$) of the turnings points can coincide with the horizons. Since region (V) intersects with region (b) the orbit crosses $\vartheta=\frac{\pi}{2}$. In the special case $\tK=(\ta E-\tL)^2$ one zero is now at $\tr=0$ so that the orbit turns into a terminating orbit and the motion takes place in the equatorial plane.
\end{enumerate}

\begin{figure}[h]
 \centering
 \subfigure[$\delta=1$, $\ta=0.45$, $\tL=0.2$, $J=1$ and $\tK=2$: Examples of the orbit types A, B, C and G.]{
   \includegraphics[width=5cm]{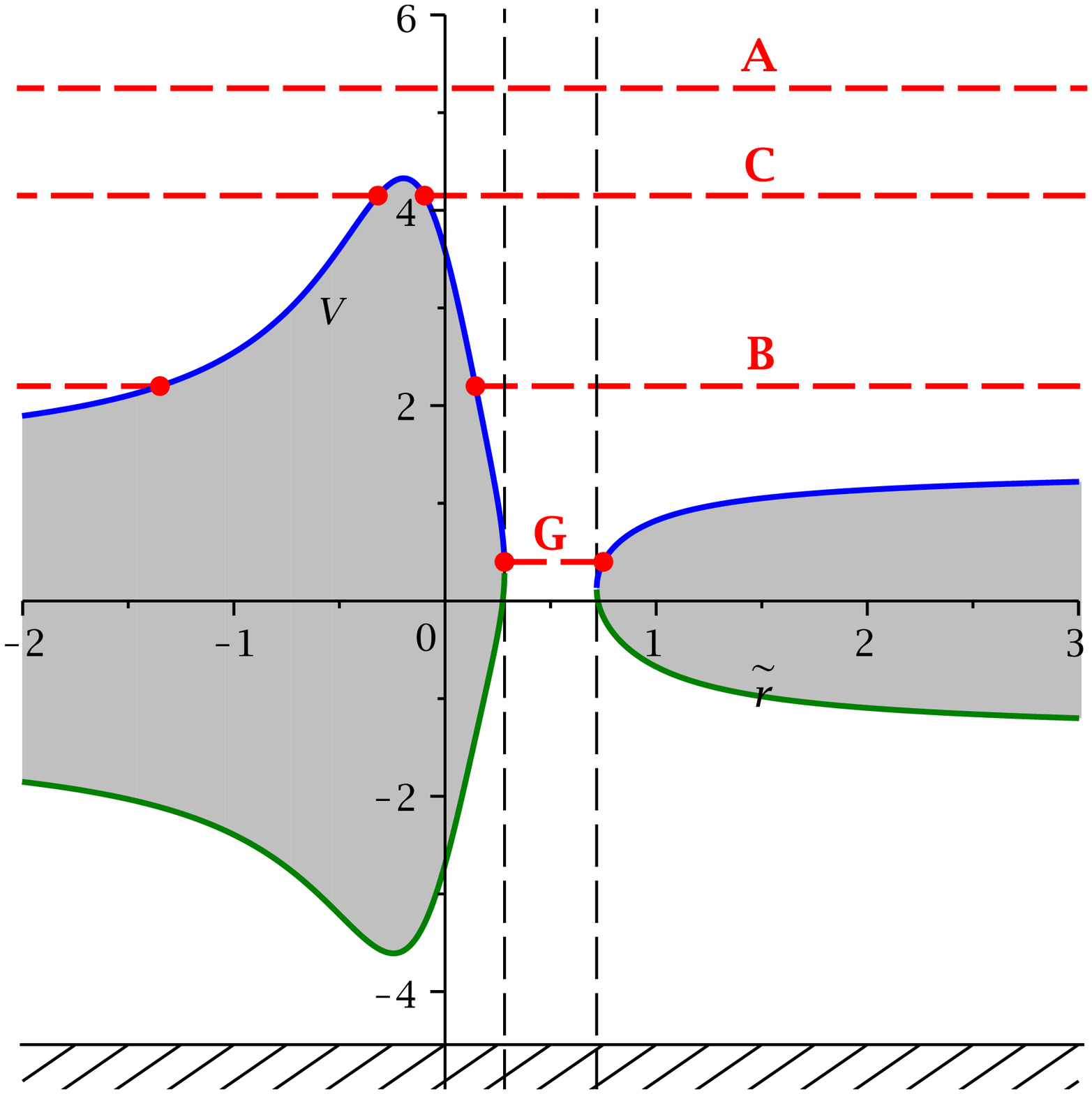}
 }
 \subfigure[$\delta=0$, $\ta=0.4$, $\tL=2$, $J=\sqrt{2}$ and $\tK=5$: Examples of the orbit types D and E. In the case of type D, here a stable bound orbit for light is possible.]{
   \includegraphics[width=5cm]{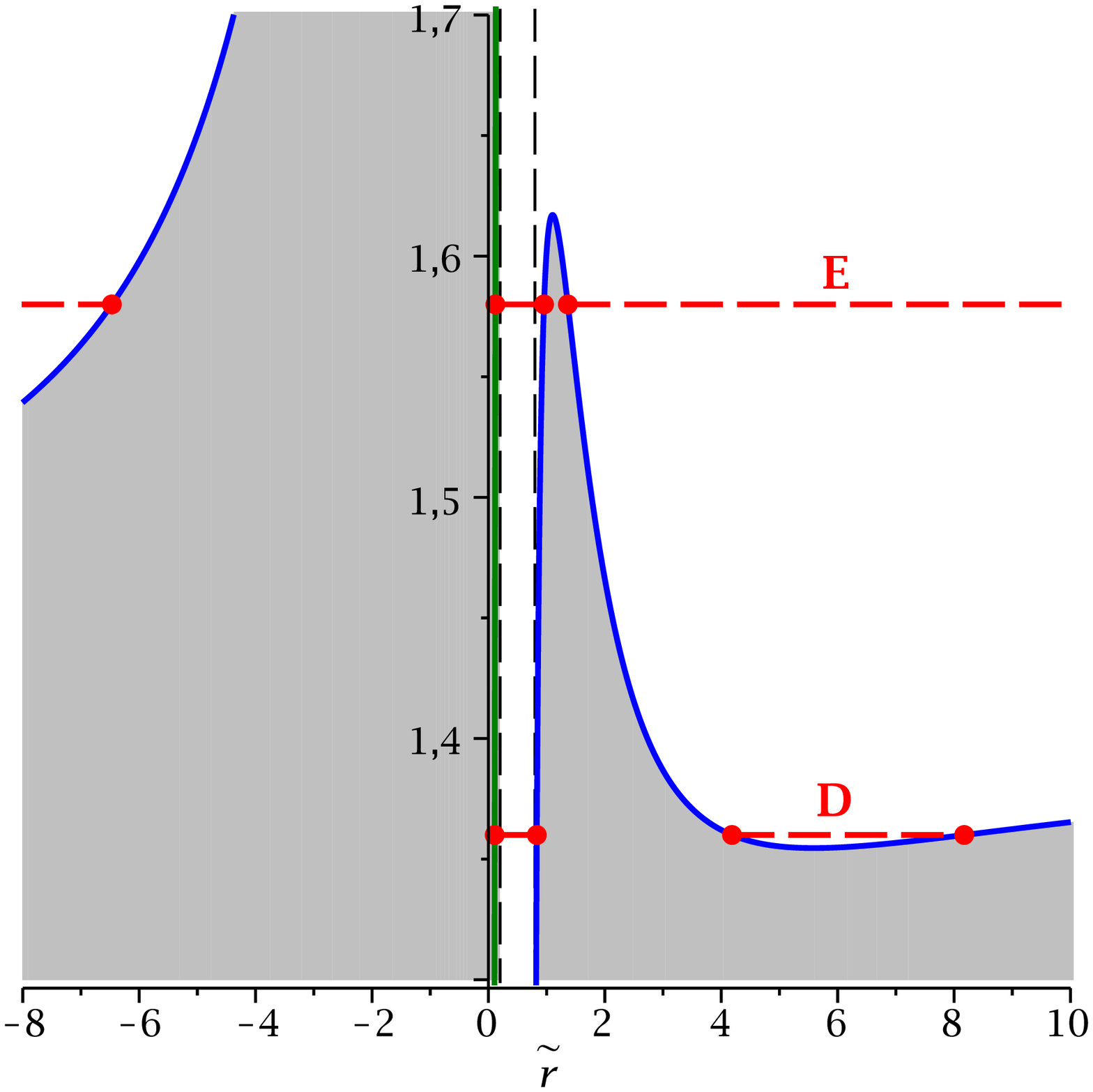}
   \label{pic:kerr-potential-b}
 }
 \subfigure[$\delta=1$, $\ta=0.45$, $\tL=1$, $J=0.2$ and $\tK=0.1$: Examples of the orbit type F.]{
   \includegraphics[width=5cm]{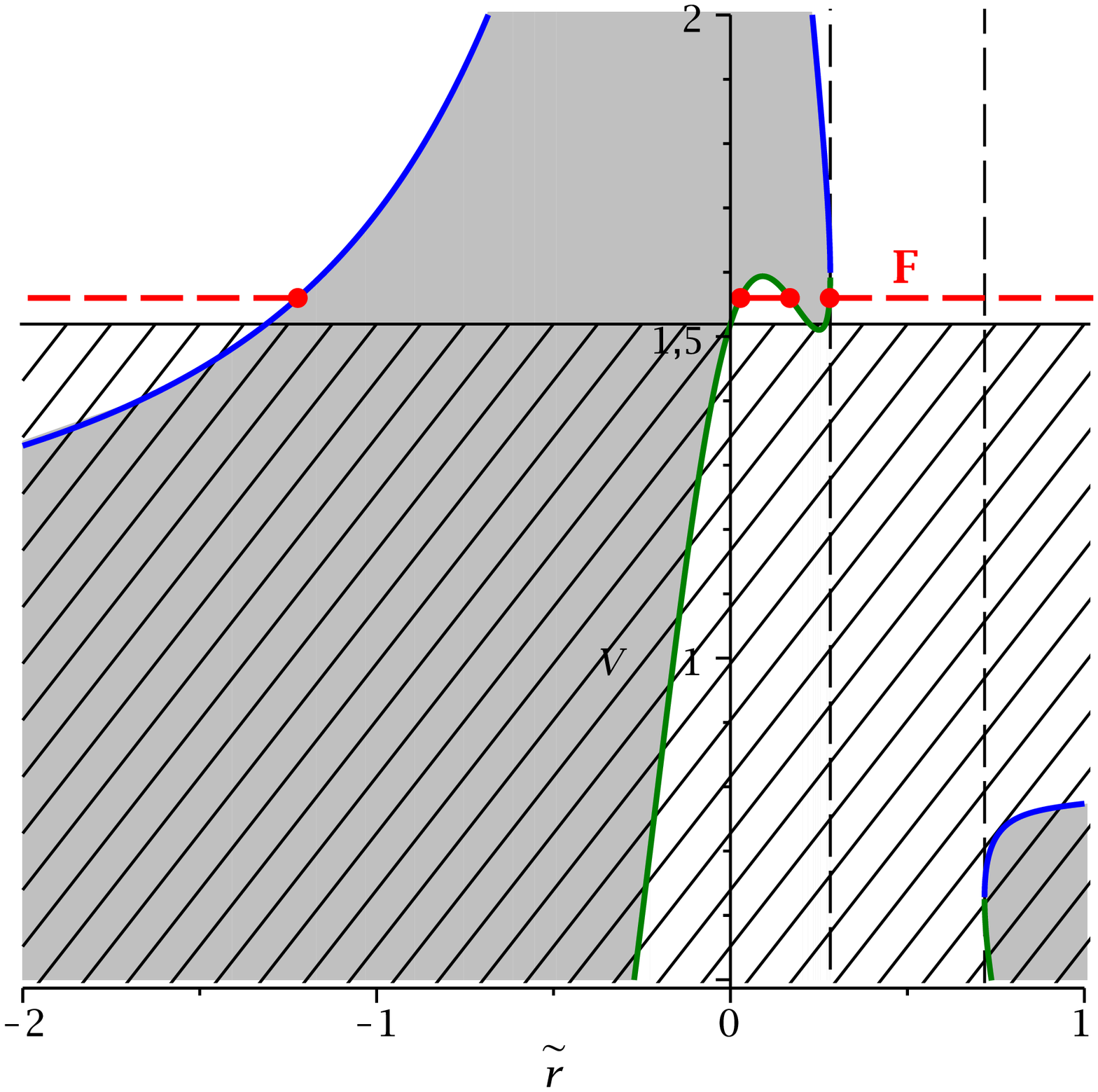}
 }
 \caption{Examples of the effective potential in the rotating black string spacetime. In grey area no motion is possible, since here $R<0$. In the dashed area the motion is forbidden by the $\vartheta$-equation. Horizontal red dashed lines represent energies and red points mark the turning points. The horizons are marked by vertical black dashed lines.}
 \label{pic:kerr-potential}
\end{figure}

Table \ref{tab:kerr-type-orbits} shows all possible orbits types of orbits in the rotating black string spacetime. Some examples of energies corressponding to the various orbit types in the effective potential can be seen in figure \ref{pic:kerr-potential}.

If we consider the original Kerr spacetime then the regions (III) and (V) are not present for $\delta=0$, but in the rotating black string spacetime all five regions are present both for $\delta=0$ and $\delta=1$. That means in the rotating black string spacetime we have bound orbits for light with $\tr>\tr_+$ (see figure \ref{pic:kerr-potential-b}) which were not possible in the Kerr spacetime.

\begin{table}[h]
\begin{center}
\begin{tabular}{|lccll|}\hline
type & zeros & region  & range of $\tr$ & orbit \\
\hline\hline
A & 0 & Id &
\begin{pspicture}(-4,-0.2)(3.5,0.2)
\psline[linewidth=0.5pt]{->}(-4,0)(3.5,0)
\psline[linewidth=0.5pt](-2.5,-0.2)(-2.5,0.2)
\psline[linewidth=0.5pt,doubleline=true](-0.5,-0.2)(-0.5,0.2)
\psline[linewidth=0.5pt,doubleline=true](1,-0.2)(1,0.2)
\psline[linewidth=1.2pt]{-}(-4,0)(3.5,0)
\end{pspicture}
  & TrO
\\  \hline
B  & 2 & IIb &
\begin{pspicture}(-4,-0.2)(3.5,0.2)
\psline[linewidth=0.5pt]{->}(-4,0)(3.5,0)
\psline[linewidth=0.5pt](-2.5,-0.2)(-2.5,0.2)
\psline[linewidth=0.5pt,doubleline=true](-0.5,-0.2)(-0.5,0.2)
\psline[linewidth=0.5pt,doubleline=true](1,-0.2)(1,0.2)
\psline[linewidth=1.2pt]{-*}(-4,0)(-3,0)
\psline[linewidth=1.2pt]{*-}(-1,0)(3.5,0)
\end{pspicture}
& EO, TEO 
\\ 
B$_-$ &  &  & 
\begin{pspicture}(-4,-0.2)(3.5,0.2)
\psline[linewidth=0.5pt]{->}(-4,0)(3.5,0)
\psline[linewidth=0.5pt](-2.5,-0.2)(-2.5,0.2)
\psline[linewidth=0.5pt,doubleline=true](-0.5,-0.2)(-0.5,0.2)
\psline[linewidth=0.5pt,doubleline=true](1,-0.2)(1,0.2)
\psline[linewidth=1.2pt]{-*}(-4,0)(-3,0)
\psline[linewidth=1.2pt]{*-}(-0.5,0)(3.5,0)
\end{pspicture}
  & EO, TEO
\\ 
B$_0$ &  &  & 
\begin{pspicture}(-4,-0.2)(3.5,0.2)
\psline[linewidth=0.5pt]{->}(-4,0)(3.5,0)
\psline[linewidth=0.5pt](-2.5,-0.2)(-2.5,0.2)
\psline[linewidth=0.5pt,doubleline=true](-0.5,-0.2)(-0.5,0.2)
\psline[linewidth=0.5pt,doubleline=true](1,-0.2)(1,0.2)
\psline[linewidth=1.2pt]{-*}(-4,0)(-3,0)
\psline[linewidth=1.2pt]{*-}(-2.5,0)(3.5,0)
\end{pspicture}
  & EO, TO
\\ \hline
C & 2 & IId &
\begin{pspicture}(-4,-0.2)(3.5,0.2)
\psline[linewidth=0.5pt]{->}(-4,0)(3.5,0)
\psline[linewidth=0.5pt](-2.5,-0.2)(-2.5,0.2)
\psline[linewidth=0.5pt,doubleline=true](-0.5,-0.2)(-0.5,0.2)
\psline[linewidth=0.5pt,doubleline=true](1,-0.2)(1,0.2)
\psline[linewidth=1.2pt]{-*}(-4,0)(-3.5,0)
\psline[linewidth=1.2pt]{*-}(-3,0)(3.5,0)
\end{pspicture}
  & EO, CTEO
\\ \hline
D & 4 & IIIb &
\begin{pspicture}(-4,-0.2)(3.5,0.2)
\psline[linewidth=0.5pt]{->}(-4,0)(3.5,0)
\psline[linewidth=0.5pt](-2.5,-0.2)(-2.5,0.2)
\psline[linewidth=0.5pt,doubleline=true](-0.5,-0.2)(-0.5,0.2)
\psline[linewidth=0.5pt,doubleline=true](1,-0.2)(1,0.2)
\psline[linewidth=1.2pt]{*-*}(-1,0)(1.5,0)
\psline[linewidth=1.2pt]{*-*}(2,0)(3,0)
\end{pspicture}
& MBO, BO 
\\
D$_\pm$ &  &  &
\begin{pspicture}(-4,-0.2)(3.5,0.2)
\psline[linewidth=0.5pt]{->}(-4,0)(3.5,0)
\psline[linewidth=0.5pt](-2.5,-0.2)(-2.5,0.2)
\psline[linewidth=0.5pt,doubleline=true](-0.5,-0.2)(-0.5,0.2)
\psline[linewidth=0.5pt,doubleline=true](1,-0.2)(1,0.2)
\psline[linewidth=1.2pt]{*-*}(-0.5,0)(1,0)
\psline[linewidth=1.2pt]{*-*}(2,0)(3,0)
\end{pspicture}
& MBO, BO 
\\
D$_-$ &  &  &
\begin{pspicture}(-4,-0.2)(3.5,0.2)
\psline[linewidth=0.5pt]{->}(-4,0)(3.5,0)
\psline[linewidth=0.5pt](-2.5,-0.2)(-2.5,0.2)
\psline[linewidth=0.5pt,doubleline=true](-0.5,-0.2)(-0.5,0.2)
\psline[linewidth=0.5pt,doubleline=true](1,-0.2)(1,0.2)
\psline[linewidth=1.2pt]{*-*}(-0.5,0)(1.5,0)
\psline[linewidth=1.2pt]{*-*}(2,0)(3,0)
\end{pspicture}
& MBO, BO 
\\
D$_+$ &  &  &
\begin{pspicture}(-4,-0.2)(3.5,0.2)
\psline[linewidth=0.5pt]{->}(-4,0)(3.5,0)
\psline[linewidth=0.5pt](-2.5,-0.2)(-2.5,0.2)
\psline[linewidth=0.5pt,doubleline=true](-0.5,-0.2)(-0.5,0.2)
\psline[linewidth=0.5pt,doubleline=true](1,-0.2)(1,0.2)
\psline[linewidth=1.2pt]{*-*}(-1,0)(1,0)
\psline[linewidth=1.2pt]{*-*}(2,0)(3,0)
\end{pspicture}
& MBO, BO 
\\ \hline
E & 4 & IVb &
\begin{pspicture}(-4,-0.2)(3.5,0.2)
\psline[linewidth=0.5pt]{->}(-4,0)(3.5,0)
\psline[linewidth=0.5pt](-2.5,-0.2)(-2.5,0.2)
\psline[linewidth=0.5pt,doubleline=true](-0.5,-0.2)(-0.5,0.2)
\psline[linewidth=0.5pt,doubleline=true](1,-0.2)(1,0.2)
\psline[linewidth=1.2pt]{-*}(-4,0)(-3,0)
\psline[linewidth=1.2pt]{*-*}(-1,0)(1.5,0)
\psline[linewidth=1.2pt]{*-}(2,0)(3.5,0)
\end{pspicture}
  & EO, MBO, EO
\\
E$_-$ &  &  &
\begin{pspicture}(-4,-0.2)(3.5,0.2)
\psline[linewidth=0.5pt]{->}(-4,0)(3.5,0)
\psline[linewidth=0.5pt](-2.5,-0.2)(-2.5,0.2)
\psline[linewidth=0.5pt,doubleline=true](-0.5,-0.2)(-0.5,0.2)
\psline[linewidth=0.5pt,doubleline=true](1,-0.2)(1,0.2)
\psline[linewidth=1.2pt]{-*}(-4,0)(-3,0)
\psline[linewidth=1.2pt]{*-*}(-0.5,0)(1.5,0)
\psline[linewidth=1.2pt]{*-}(2,0)(3.5,0)
\end{pspicture}
  & EO, MBO, EO
\\
E$_+$ &  &  &
\begin{pspicture}(-4,-0.2)(3.5,0.2)
\psline[linewidth=0.5pt]{->}(-4,0)(3.5,0)
\psline[linewidth=0.5pt](-2.5,-0.2)(-2.5,0.2)
\psline[linewidth=0.5pt,doubleline=true](-0.5,-0.2)(-0.5,0.2)
\psline[linewidth=0.5pt,doubleline=true](1,-0.2)(1,0.2)
\psline[linewidth=1.2pt]{-*}(-4,0)(-3,0)
\psline[linewidth=1.2pt]{*-*}(-1,0)(1,0)
\psline[linewidth=1.2pt]{*-}(2,0)(3.5,0)
\end{pspicture}
  & EO, MBO, EO
\\ \hline
F & 4 & IVb &
\begin{pspicture}(-4,-0.2)(3.5,0.2)
\psline[linewidth=0.5pt]{->}(-4,0)(3.5,0)
\psline[linewidth=0.5pt](-2.5,-0.2)(-2.5,0.2)
\psline[linewidth=0.5pt,doubleline=true](-0.5,-0.2)(-0.5,0.2)
\psline[linewidth=0.5pt,doubleline=true](1,-0.2)(1,0.2)
\psline[linewidth=1.2pt]{-*}(-4,0)(-3,0)
\psline[linewidth=1.2pt]{*-*}(-2,0)(-1,0)
\psline[linewidth=1.2pt]{*-}(2,0)(3.5,0)
\end{pspicture}
  & EO, BO, EO
\\
F$_+$ &  &  &
\begin{pspicture}(-4,-0.2)(3.5,0.2)
\psline[linewidth=0.5pt]{->}(-4,0)(3.5,0)
\psline[linewidth=0.5pt](-2.5,-0.2)(-2.5,0.2)
\psline[linewidth=0.5pt,doubleline=true](-0.5,-0.2)(-0.5,0.2)
\psline[linewidth=0.5pt,doubleline=true](1,-0.2)(1,0.2)
\psline[linewidth=1.2pt]{-*}(-4,0)(-3,0)
\psline[linewidth=1.2pt]{*-*}(-2,0)(-1,0)
\psline[linewidth=1.2pt]{*-}(1,0)(3.5,0)
\end{pspicture}
  & EO, BO, EO
\\
F$_0$ &  &  &
\begin{pspicture}(-4,-0.2)(3.5,0.2)
\psline[linewidth=0.5pt]{->}(-4,0)(3.5,0)
\psline[linewidth=0.5pt](-2.5,-0.2)(-2.5,0.2)
\psline[linewidth=0.5pt,doubleline=true](-0.5,-0.2)(-0.5,0.2)
\psline[linewidth=0.5pt,doubleline=true](1,-0.2)(1,0.2)
\psline[linewidth=1.2pt]{-*}(-4,0)(-3,0)
\psline[linewidth=1.2pt]{*-*}(-2.5,0)(-1,0)
\psline[linewidth=1.2pt]{*-}(2,0)(3.5,0)
\end{pspicture}
  & EO, TO, EO
\\
F$_{0+}$ &  &  &
\begin{pspicture}(-4,-0.2)(3.5,0.2)
\psline[linewidth=0.5pt]{->}(-4,0)(3.5,0)
\psline[linewidth=0.5pt](-2.5,-0.2)(-2.5,0.2)
\psline[linewidth=0.5pt,doubleline=true](-0.5,-0.2)(-0.5,0.2)
\psline[linewidth=0.5pt,doubleline=true](1,-0.2)(1,0.2)
\psline[linewidth=1.2pt]{-*}(-4,0)(-3,0)
\psline[linewidth=1.2pt]{*-*}(-2.5,0)(-1,0)
\psline[linewidth=1.2pt]{*-}(1,0)(3.5,0)
\end{pspicture}
  & EO, TO, EO
\\ \hline
G & 2 & Vb &
\begin{pspicture}(-4,-0.2)(3.5,0.2)
\psline[linewidth=0.5pt]{->}(-4,0)(3.5,0)
\psline[linewidth=0.5pt](-2.5,-0.2)(-2.5,0.2)
\psline[linewidth=0.5pt,doubleline=true](-0.5,-0.2)(-0.5,0.2)
\psline[linewidth=0.5pt,doubleline=true](1,-0.2)(1,0.2)
\psline[linewidth=1.2pt]{*-*}(-1,0)(1.5,0)
\end{pspicture}
& MBO 
\\
G$_\pm$ &  &  &
\begin{pspicture}(-4,-0.2)(3.5,0.2)
\psline[linewidth=0.5pt]{->}(-4,0)(3.5,0)
\psline[linewidth=0.5pt](-2.5,-0.2)(-2.5,0.2)
\psline[linewidth=0.5pt,doubleline=true](-0.5,-0.2)(-0.5,0.2)
\psline[linewidth=0.5pt,doubleline=true](1,-0.2)(1,0.2)
\psline[linewidth=1.2pt]{*-*}(-0.5,0)(1,0)
\end{pspicture}
& MBO
\\
G$_-$ &  &  &
\begin{pspicture}(-4,-0.2)(3.5,0.2)
\psline[linewidth=0.5pt]{->}(-4,0)(3.5,0)
\psline[linewidth=0.5pt](-2.5,-0.2)(-2.5,0.2)
\psline[linewidth=0.5pt,doubleline=true](-0.5,-0.2)(-0.5,0.2)
\psline[linewidth=0.5pt,doubleline=true](1,-0.2)(1,0.2)
\psline[linewidth=1.2pt]{*-*}(-0.5,0)(1.5,0)
\end{pspicture}
& MBO
\\
G$_+$ &  &  &
\begin{pspicture}(-4,-0.2)(3.5,0.2)
\psline[linewidth=0.5pt]{->}(-4,0)(3.5,0)
\psline[linewidth=0.5pt](-2.5,-0.2)(-2.5,0.2)
\psline[linewidth=0.5pt,doubleline=true](-0.5,-0.2)(-0.5,0.2)
\psline[linewidth=0.5pt,doubleline=true](1,-0.2)(1,0.2)
\psline[linewidth=1.2pt]{*-*}(-1,0)(1,0)
\end{pspicture}
& MBO
\\
G$_0$ &  &  &
\begin{pspicture}(-4,-0.2)(3.5,0.2)
\psline[linewidth=0.5pt]{->}(-4,0)(3.5,0)
\psline[linewidth=0.5pt](-2.5,-0.2)(-2.5,0.2)
\psline[linewidth=0.5pt,doubleline=true](-0.5,-0.2)(-0.5,0.2)
\psline[linewidth=0.5pt,doubleline=true](1,-0.2)(1,0.2)
\psline[linewidth=1.2pt]{*-*}(-2.5,0)(1.5,0)
\end{pspicture}
& TO
\\
G$_{0+}$ &  &  &
\begin{pspicture}(-4,-0.2)(3.5,0.2)
\psline[linewidth=0.5pt]{->}(-4,0)(3.5,0)
\psline[linewidth=0.5pt](-2.5,-0.2)(-2.5,0.2)
\psline[linewidth=0.5pt,doubleline=true](-0.5,-0.2)(-0.5,0.2)
\psline[linewidth=0.5pt,doubleline=true](1,-0.2)(1,0.2)
\psline[linewidth=1.2pt]{*-*}(-2.5,0)(1,0)
\end{pspicture}
& TO
\\ \hline\hline
\end{tabular}
\caption{Types of orbits of light and particles in the rotating black string spacetime. The thick lines represent the range of the orbits. The turning points are shown by thick dots. The horizons are indicated by a vertical double line. The single vertical line represents $\tr=0$.}
\label{tab:kerr-type-orbits}
\end{center}
\end{table}

\subsection{Solution of the geodesic equations}

In this section we present the analytical solution of the equations of motion \eqref{eqn:kerr-r-equation}-\eqref{eqn:kerr-t-equation}.

\subsubsection{The $\tr$-equation}

The right hand side of the $\tr$-equation \eqref{eqn:kerr-r-equation} is a polynomial of fourth order with the coefficients:
\begin{eqnarray}
 a_4 &=& E^2-(\delta+J^2)\\
 a_3 &=& \delta+J^2\\
 a_2 &=& 2\ta E(\ta E-\tL)-\ta^2(\delta+J^2)-\tK\\
 a_1 &=& \tK\\
 a_0 &=& \ta^2[(\ta E-\tL)^2-\tK]
\end{eqnarray}
The polynomial $R=\sum _{i=1}^4 a_i\tr^i$ can be reduced to cubic order by the substution $\tr=\pm\frac{1}{x}+\tr_R$ (where $\tr_R$ is a zero of $R$): $R'= \sum _{i=0}^3 b_i x^i$. A further substitution $x=\frac{1}{b_3}\left( 4y-\frac{b_2}{3}\right)$ transforms $R'$ into the Weierstra{\ss} form so that equation \eqref{eqn:kerr-r-equation} turns into
\begin{equation}
\left(\frac{dy}{d\gamma}\right)^2=4y^3-g_2^{\tr}y-g_3^{\tr}= P_3^{\tr} (y) \, ,
\label{eqn:weierstrass2}
\end{equation}
where 
\begin{equation}
g_2^{\tr}=\frac{b_2^2}{12} - \frac{b_1b_3}{4} \, , \qquad  g_3^{\tr}=\frac{b_1b_2b_3}{48} - \frac{b_0b_3^2}{16}-\frac{b_2^3}{216} \ .
\end{equation}
The differential equation \eqref{eqn:weierstrass2} is of elliptic type and is solved by the Weierstra{\ss} $\wp$-function \cite{Markushevich:1967}
\begin{equation}
y(\gamma) = \wp\left(\gamma - \gamma'_{\rm in}; g_2^{\tr}, g_3^{\tr}\right) \ ,
\end{equation}
where $\gamma'_{\rm in}=\gamma_{\rm in}+\int^\infty_{y_{\rm in}}{\frac{dy}{\sqrt{4y^3-g_2^{\tr}y-g_3^{\tr}}}}$
with $y_{\rm in}=\pm\frac{b_3}{4\tr_{\rm in}} + \frac{b_2}{12}$.
Then the solution of \eqref{eqn:kerr-r-equation} acquires the form
\begin{equation}
\tr=\pm \frac{b_3}{4 \wp\left(\gamma - \gamma'_{\rm in}; g_2^{\tr}, g_3^{\tr}\right) - \frac{b_2}{3}} +\tr_R\ .
\end{equation}

\subsubsection{The $\vartheta$-equation}

To solve the $\vartheta$-equation \eqref{eqn:kerr-theta-equation} we substitute $\nu=\cos^2\vartheta$ (with $\nu\in[0,1]$) in equation \eqref{eqn:kerr-theta-equation}:
\begin{equation}
 \left( \frac{\dd\nu}{\dd\gamma} \right)^2 = 4\nu(1-\nu)[\tK-(\delta+J^2)\ta\nu]-4\nu[\ta E/1-\nu)-\tL]^2
\label{eqn:kerr-theta-poly}
\end{equation}
The right hand side of equation \eqref{eqn:kerr-theta-poly} is a polynomial of third order $\sum^3_{i=1}c_i\nu^i$ with the coefficients:
\begin{eqnarray}
 c_3 &=& -4\ta^2[E^2-(\delta+J^2)]\\
 c_2 &=& 4[2\ta E(\ta E-\tL)-\ta^2(\delta+J^2)-\tK]\\
 c_1 &=& 4[\tK-(\ta E-\tK)^2] \, .
\end{eqnarray}
So equation \eqref{eqn:kerr-theta-poly} can be transformed into the Weierstra{\ss} form using the substitution $\nu=\frac{1}{c_3}\left(4u-\frac{c_2}{3}\right)$:
\begin{equation}
 \left( \frac{\dd u}{\dd\gamma} \right)^2 =4u^3-g_2^\vartheta u -g_3^\vartheta = P_3^{\vartheta} (u) \, ,
\label{eqn:weierstrass3}
\end{equation}
where
\begin{equation}
g_2^\vartheta=\frac{c_2^2}{12} - \frac{c_1c_3}{4} \, , \qquad  g_3^\vartheta=\frac{c_1c_2c_3}{48}-\frac{c_2^3}{216} \ .
\end{equation}
Equation \eqref{eqn:weierstrass3} is solved by the Weierstra{\ss} $\wp$-function and the solution $\vartheta(\gamma)$ of equation \eqref{eqn:kerr-theta-equation} can then be obtained by resubstitution:
\begin{equation}
 \vartheta(\gamma) = \arccos\left( \pm \sqrt{\frac{1}{c_3}\left( 4\wp(\gamma-\gamma''_{\rm in};g_2^\vartheta,g_3^\vartheta) -\frac{c_2}{3} \right)} \right)
\end{equation}
with $\gamma''_{\rm in}=\gamma_{\rm in}+\int^\infty_{u_{\rm in}}{\frac{du}{\sqrt{4u^3-g_2^\vartheta y-g_3^\vartheta}}}$ and $u_{\rm in}=\frac{c_3}{4}\cos^2\vartheta_{\rm in} + \frac{c_2}{12}$.

\subsubsection{The $\varphi$-equation}
\label{sec:kerr-phisol}

Using the $\tr$-equation \eqref{eqn:kerr-r-equation} and the $\vartheta$-equation \eqref{eqn:kerr-theta-equation}, we can write the $\varphi$-equation \eqref{eqn:kerr-phi-equation} in the following way:
\begin{equation}
  \dd \varphi = \frac{\ta}{\tDr}[(\tr^2+\ta^2)E-\ta\tL]\frac{\dd\tr}{\sqrt{R}} - \frac{1}{\sin^2\vartheta}(\ta E \sin^2\vartheta -\tL)\frac{\dd\vartheta}{\sqrt{\Theta}} \, .
\end{equation}
So the $\varphi$-equation consists of a $\tr$-dependent integral and a $\vartheta$-dependent integral:
\begin{equation}
 \varphi - \varphi_{\rm in} = \int_{\tr_{\rm in}}^{\tr} \! \frac{\ta}{\tDr}[(\tr'^2+\ta^2)E-\ta\tL]\frac{\dd\tr'}{\sqrt{R}} - \int_{\vartheta_{\rm in}}^{\vartheta} \! \frac{1}{\sin^2\vartheta'}(\ta E \sin^2\vartheta' -\tL)\frac{\dd\vartheta'}{\sqrt{\Theta}} = I_{\tr} - I_\vartheta \, .
\end{equation}

Let us first consider $I_{\tr}$. Here we substitute $\tr=\pm\frac{b_3}{4y-\frac{b_2}{3}}+\tr_R$ and apply a partial fraction decomposition, so that $I_{\tr}$ turns into
\begin{equation}
 I_{\tr}= \int_{y_{\rm in}}^{y} \! C_0 + \sum^2_{i=1}\frac{C_i}{y'-p_i} \frac{\dd y'}{\sqrt{P^{\tr}_3(y')}} \, ,
\end{equation}
where $p_1=\frac{b_2(\tr_+-\tr_R)\pm b_3}{12(\tr_+-\tr_R)}$ and $p_2=\frac{b_2(\tr_--\tr_R)\pm b_3}{12(\tr_--\tr_R)}$ are first order poles of $I_{\tr}$. The sign of $\pm b_3$ depends on the chosen sign in the substitution $\tr=\pm\frac{b_3}{4y-\frac{b_2}{3}}+\tr_R$. $C_i$ are constants that arise from the partial fraction decomposition and depend on the parameters of the metric and the test particle.
Now we substitute $y=\wp\left(\gamma - \gamma'_{\rm in}; g_2^{\tr}, g_3^{\tr}\right)=:\wp_{\tr}(v)$ with $v=\gamma-\gamma_{\rm in}'$:
\begin{equation}
 I_{\tr}= \int_{v_{\rm in}}^{v} \! C_0 + \sum^2_{i=1}\frac{C_i}{\wp_{\tr}(v')-p_i} \dd v' \, ,
\end{equation}

The integral $I_\vartheta$ can be transformed in the same way by substituting first $\nu=\cos^2\vartheta$, then $\nu=\frac{1}{c_3}\left(4u-\frac{c_2}{3}\right)$ and finally $u=\wp(\gamma-\gamma''_{\rm in};g_2^\vartheta,g_3^\vartheta)=:\wp_{\vartheta}(\tv)$ with $\tv=\gamma-\gamma_{\rm in}''$:
\begin{equation}
 I_{\vartheta}= \int_{\tv_{\rm in}}^{\tv} \! \ta E +\frac{c_3\tL}{4}\frac{1}{\wp_{\vartheta}(\tv')-q} \dd \tv' \, ,
\end{equation}
where $q=\frac{c_3}{4}+\frac{c_2}{12}$.

The integrals $I_{\tr}$ and $I_\vartheta$ are of elliptic type and can be solved  in terms of the elliptic $\wp$-, $\sigma$- and $\zeta$-function as shown in \cite{Kagramanova:2010bk, Grunau:2010gd}. Then the final solution of the $\varphi$-equation \eqref{eqn:kerr-phi-equation} is
\begin{equation}
 \begin{split}
  \varphi(\gamma) &= C_0(v-v_{\rm in}) + \sum^2_{i=1}\frac{C_i}{\wp'_{\tr}(v_i)}\left( 2\zeta_{\tr}(v_i)(v-v_{\rm in}) + \ln\frac{\sigma_{\tr}(v-v_i)}{\sigma_{\tr}(v_{\rm in}-v_i)} - \ln\frac{\sigma_{\tr}(v+v_i)}{\sigma_{\tr}(v_{\rm in}+v_i)}\right) \\
 &-\ta E(\tv-\tv_{\rm in}) - \frac{c_3\tL}{4\wp'_{\vartheta}(\tv_q)}\left( 2\zeta_{\vartheta}(\tv_q)(\tv-\tv_{\rm in}) + \ln\frac{\sigma_{\vartheta}(\tv-\tv_q)}{\sigma_{\vartheta}(\tv_{\rm in}-v_q)} - \ln\frac{\sigma_{\vartheta}(\tv+\tv_q)}{\sigma_{\vartheta}(\tv_{\rm in}+\tv_q)}\right) + \varphi_{\rm in}
 \end{split}
\end{equation}
where $p_i=\wp_{\tr}(v_i)$, $q=\wp_{\vartheta}(\tv_q)$, $v=\gamma-\gamma_{\rm in}'$,  $\tv=\gamma-\gamma_{\rm in}''$  and
\begin{eqnarray}
\wp_{\tr}(v) &= \wp (v, g_2^{\tr}, g_3^{\tr})\, , \qquad \wp_\vartheta (\tv)&= \wp (\tv, g_2^{\vartheta}, g_3^{\vartheta}) \, ,\nonumber\\
\zeta_{\tr}(v) &= \zeta (v, g_2^{\tr}, g_3^{\tr})\, , \qquad \zeta_\vartheta (\tv)&= \zeta (\tv, g_2^{\vartheta}, g_3^{\vartheta}) \, ,\\
\sigma_{\tr}(v) &= \sigma (v, g_2^{\tr}, g_3^{\tr})\, , \qquad \sigma_\vartheta (\tv)&= \sigma (\tv, g_2^{\vartheta}, g_3^{\vartheta}) \, .\nonumber
\end{eqnarray}

\subsubsection{The $\tw$-equation}

Using the $\tr$-equation \eqref{eqn:kerr-r-equation} and the $\vartheta$-equation \eqref{eqn:kerr-theta-equation}, we can write the $\tw$-equation \eqref{eqn:kerr-w-equation} in the following way:
\begin{equation}
 \dd\tw = J\trho^2 \dd \gamma = J\tr^2\frac{\dd\tr}{\sqrt{R}} + J\ta^2\cos^2\vartheta\frac{\dd\vartheta}{\sqrt{\Theta}} \, .
\label{eqn:wsplit}
\end{equation}
Like the $\varphi$-equation, the $\tw$-equation consist of a $\tr$-dependent part and a $\vartheta$-dependent part. We integrate \eqref{eqn:wsplit} and use the same substitutions as in the previous section \ref{sec:kerr-phisol}:
\begin{equation}
 \tw - \tw_{\rm in} = J \int_{v_{\rm in}}^{v} \!\left( \tr_R^2 \pm \frac{b_3\tr_R}{2}\frac{1}{\wp(v')-p} + \frac{b_3^2}{16}\frac{1}{(\wp(v')-p)^2} \right) \dd v' + J\ta^2 \int_{\tv_{\rm in}}^{\tv} \! \left( \frac{4}{c_3} \wp_{\vartheta}(\tv) - \frac{c_2}{3c_3} \right) \dd \tv' \, ,
\end{equation}
where $p=\frac{b_2}{12}$.
The occurring elliptic intregrals of the third kind can be solved as shown in \cite{Kagramanova:2010bk, Grunau:2010gd}. Furthermore we use the relation $\int \wp (v) \dd v= - \zeta(v)$. The final solution of the $\tw$-equation is
\begin{equation}
 \begin{split}
  \tw(\gamma) &= J\tr_R^2(v-v_{\rm in}) + \left( \pm \frac{Jb_3\tr_R}{2\wp'_{\tr}(v_p)} - \frac{Jb_3^2\wp_{\tr}''(v_p)}{16(\wp_{\tr}'(v_p))^3} \right) \left( 2\zeta_{\tr}(v_p)(v-v_{\rm in}) + \ln\frac{\sigma_{\tr}(v-v_p)}{\sigma_{\tr}(v_{\rm in}-v_p)} - \ln\frac{\sigma_{\tr}(v+v_p)}{\sigma_{\tr}(v_{\rm in}+v_p)}\right) \\
 &-\frac{Jb_3^2}{16}\frac{1}{(\wp_{\tr}'(v_p))^2}\left(2\wp_{\tr}(v_p)(v-v_{\rm in}) + 2(\zeta_{\tr}(v)-\zeta_{\tr}(v_{\rm in})) + \frac{\wp_{\tr}'(v)}{\wp_{\tr}(v)-\wp_{\tr}(v_p)} - \frac{\wp_{\tr}'(v_{\rm in})}{\wp_{\tr}(v_{\rm in})-\wp_{\tr}(v_p)}\right) \\
 & -\frac{4J\ta^2}{c_3}(\zeta_{\vartheta}(\tv)-\zeta_{\vartheta}(\tv_{\rm in})) - \frac{J\ta^2c_2}{3c_3}(\tv-\tv_{\rm in})+ \tw_{\rm in} \, ,
 \end{split}
\end{equation}
where $p=\wp_{\tr}(v_p)$.

\subsubsection{The $\tlt$-equation}

Using the $\tr$-equation \eqref{eqn:kerr-r-equation} and the $\vartheta$-equation \eqref{eqn:kerr-theta-equation}, we can write the $\tlt$-equation \eqref{eqn:kerr-t-equation} in the following way:
\begin{equation}
\dd \tlt = \frac{\tr^2+\ta^2}{\tDr}[(\tr^2+\ta^2)E-\ta\tL] \frac{\dd\tr}{\sqrt{R}} - \ta(\ta E \sin^2\vartheta -\tL) \frac{\dd\vartheta}{\sqrt{\Theta}} \, .
\label{eqn:tsplit}
\end{equation}
Like the $\varphi$-equation and the $\tw$-equation, the $\tlt$-equation consist of a $\tr$-dependent part and a $\vartheta$-dependent part. We integrate \eqref{eqn:tsplit} and use the same substitutions as in section \ref{sec:kerr-phisol}:
\begin{equation}
 \tlt - \tlt_{\rm in} = \int_{v_{\rm in}}^{v} \! \left( C'_0 + \sum^2_{i=1}\frac{C'_i}{\wp_{\tr}(v')-p_i} + \frac{C'_3}{(\wp_{\tr}(v')-p_3)^2} \right) \dd v' - \int_{\tv_{\rm in}}^{\tv} \! \left( \ta^2E-\ta\tL+\frac{c_2}{3} - \frac{4\ta^2E}{c_3}\wp_{\vartheta}(\tv) \right) \dd \tv' \, ,
\end{equation}
where  $p_1=\frac{b_2(\tr_+-\tr_R)\pm b_3}{12(\tr_+-\tr_R)}$ and $p_2=\frac{b_2(\tr_--\tr_R)\pm b_3}{12(\tr_--\tr_R)}$ are first order poles and $p_3=\frac{b_2}{12}$ is a second order pole. $C'_i$ are constants arising from a partical fraction decomposition of the $\tr$-dependent part.
The occurring elliptic intregrals of the third kind can be solved as shown in \cite{Kagramanova:2010bk, Grunau:2010gd}. Furthermore we use the relation $\int \wp (v) \dd v= - \zeta(v)$. The final solution of the $\tlt$-equation is
\begin{equation}
 \begin{split}
  \tlt(\gamma) &= C'_0(v-v_{\rm in}) + \sum^2_{i=1}\frac{C'_i}{\wp'_{\tr}(v_i)}\left( 2\zeta_{\tr}(v_i)(v-v_{\rm in}) + \ln\frac{\sigma_{\tr}(v-v_i)}{\sigma_{\tr}(v_{\rm in}-v_i)} - \ln\frac{\sigma_{\tr}(v+v_i)}{\sigma_{\tr}(v_{\rm in}+v_i)}\right) \\
 & -C'_3\frac{\wp''_{\tr}(v_3)}{(\wp'_{\tr}(v_3))^2}\left( 2\zeta_{\tr}(v_3)(v-v_{\rm in}) + \ln\frac{\sigma_{\tr}(v-v_3)}{\sigma_{\tr}(v_{\rm in}-v_3)} - \ln\frac{\sigma_{\tr}(v+v_3)}{\sigma_{\tr}(v_{\rm in}+v_3)}\right) \\
 &-\frac{C'_3}{(\wp_{\tr}'(v_3))^2}\left(2\wp_{\tr}(v_3)(v-v_{\rm in}) + 2(\zeta_{\tr}(v)-\zeta_{\tr}(v_{\rm in})) + \frac{\wp_{\tr}'(v)}{\wp_{\tr}(v)-\wp_{\tr}(v_3)} - \frac{\wp_{\tr}'(v_{\rm in})}{\wp_{\tr}(v_{\rm in})-\wp_{\tr}(v_3)}\right) \\
 & -\left( \ta^2E-\ta\tL+\frac{c_2}{3}\right) (\tv-\tv_{\rm in}) -  \frac{4\ta^2E}{c_3}(\zeta_{\vartheta}(\tv)-\zeta_{\vartheta}(\tv_{\rm in}))+ \tlt_{\rm in} \, ,
 \end{split}
\end{equation}
where $p_i=\wp_{\tr}(v_i)$.

\subsection{The orbits}

With these analytical results we have found the complete set of orbits for light and test particles in the rotating black string spacetime. Depending on the parameters $\delta$, $\tL$, $J$, $\ta$ and $E$ various orbits are possible.
Note that the horizons in the Kerr spacetime and in the rotating black string spacetime ($x$-$y$-$z$-plots) are not spheres but ellipsoids. A $x$-$y$-$w$-plot of the horizons in the rotating black string spacetime will show cylinders.

\begin{figure}[ht]
 \centering
 \subfigure[$x$-$y$-$z$-plot]{
   \includegraphics[width=6cm]{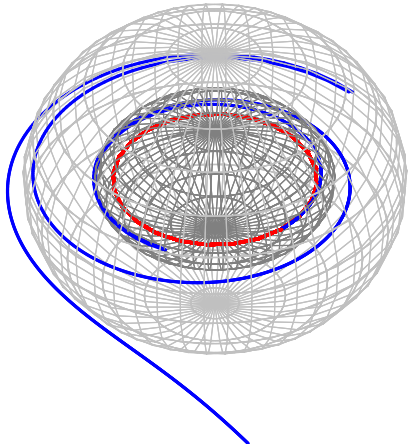}
 }\qquad\qquad
 \subfigure[$x$-$y$-$w$-plot]{
   \includegraphics[width=6cm]{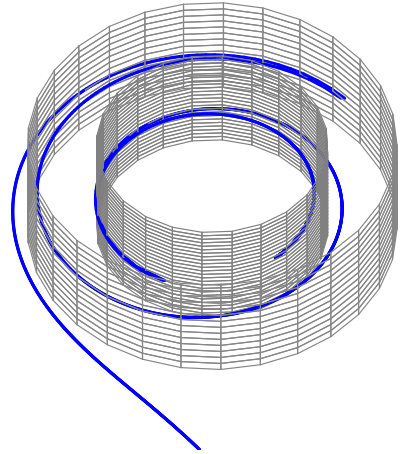}
 }
 \caption{$\delta=1$, $\ta=0.45$, $\tL=0.8$, $J=0.2$, $\tK=(\ta E-\tL)^2=2.1025$ and $E=5$:\newline
          Terminating orbit for particles in the rotating black string spacetime. The ellipsoids or cylinders are the horizons. In the left picture, the position of the ring singularity is marked by a red circle.}
 \label{pic:kerr-to-particle}
\end{figure}

\begin{figure}[ht]
 \centering
 \subfigure[$x$-$y$-$z$-plot]{
   \includegraphics[width=6cm]{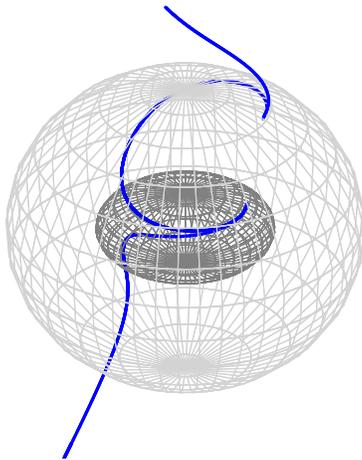}
 }\qquad\qquad
 \subfigure[$x$-$y$-$w$-plot]{
   \includegraphics[width=6cm]{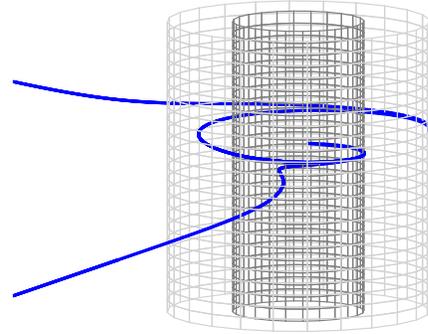}
 }
 \caption{$\delta=1$, $\ta=0.4$, $\tL=0.6$, $J=1.5$, $\tK=1$ and $E=5$:\newline
          Transit orbit for particles in the rotating black string spacetime. The ellipsoids or cylinders are the horizons. In the left picture, the position of the ring singularity is marked by a red circle.}
 \label{pic:kerr-tro-particle}
\end{figure}

Figure \ref{pic:kerr-to-particle} shows a terminating orbit in the rotating black string spacetime, in the $x$-$y$-$z$-plot this orbit lies entirely in the equatorial plane. To hit the ring singularity at $\trho^2=0$ the parameters corresponding to the orbit have to fullfill the condition  $\tK=(\ta E-\tL)^2$. Otherwise it is possible for a geodesic to cross $\tr=0$ without touching the singularity. Like in the Kerr spacetime the radial coordinate $\tr$ can take negative values. A transit orbit wich starts at positve $\tr$, crosses $\tr=0$ and the continues at negative $\tr$ can be seen in figure \ref{pic:kerr-tro-particle}. When $\tr$ changes from positive to negative it looks like the particle is reflected, which can be interpreted as gravity becoming repulsive for negative $\tr$-values (see \cite{ONeill:1995,Carter:1968rr,Carter:1966zza}).

\begin{figure}[ht]
 \centering
 \subfigure[$x$-$y$-$z$-plot]{
   \includegraphics[width=6cm]{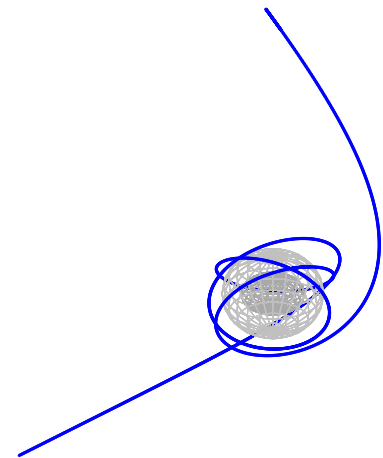}
 }\qquad\qquad
 \subfigure[$x$-$y$-$w$-plot]{
   \includegraphics[width=6cm]{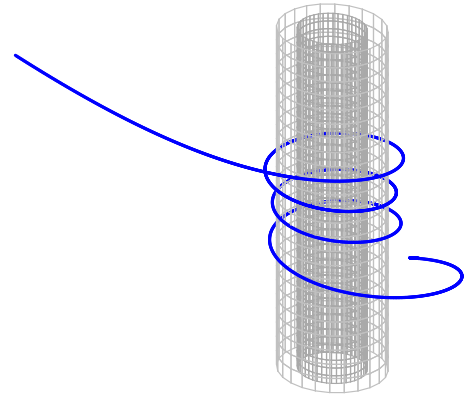}
 }
 \caption{$\delta=1$, $\ta=0.45$, $\tL=1.5$, $J=0.6$, $\tK=2$ and $E=1.251$:\newline
          Escape orbit for particles in the rotating black string spacetime. The ellipsoids or cylinders are the horizons.}
 \label{pic:kerr-eo-particle}
\end{figure}

\begin{figure}[ht]
 \centering
 \subfigure[$x$-$y$-$z$-plot]{
   \includegraphics[width=6cm]{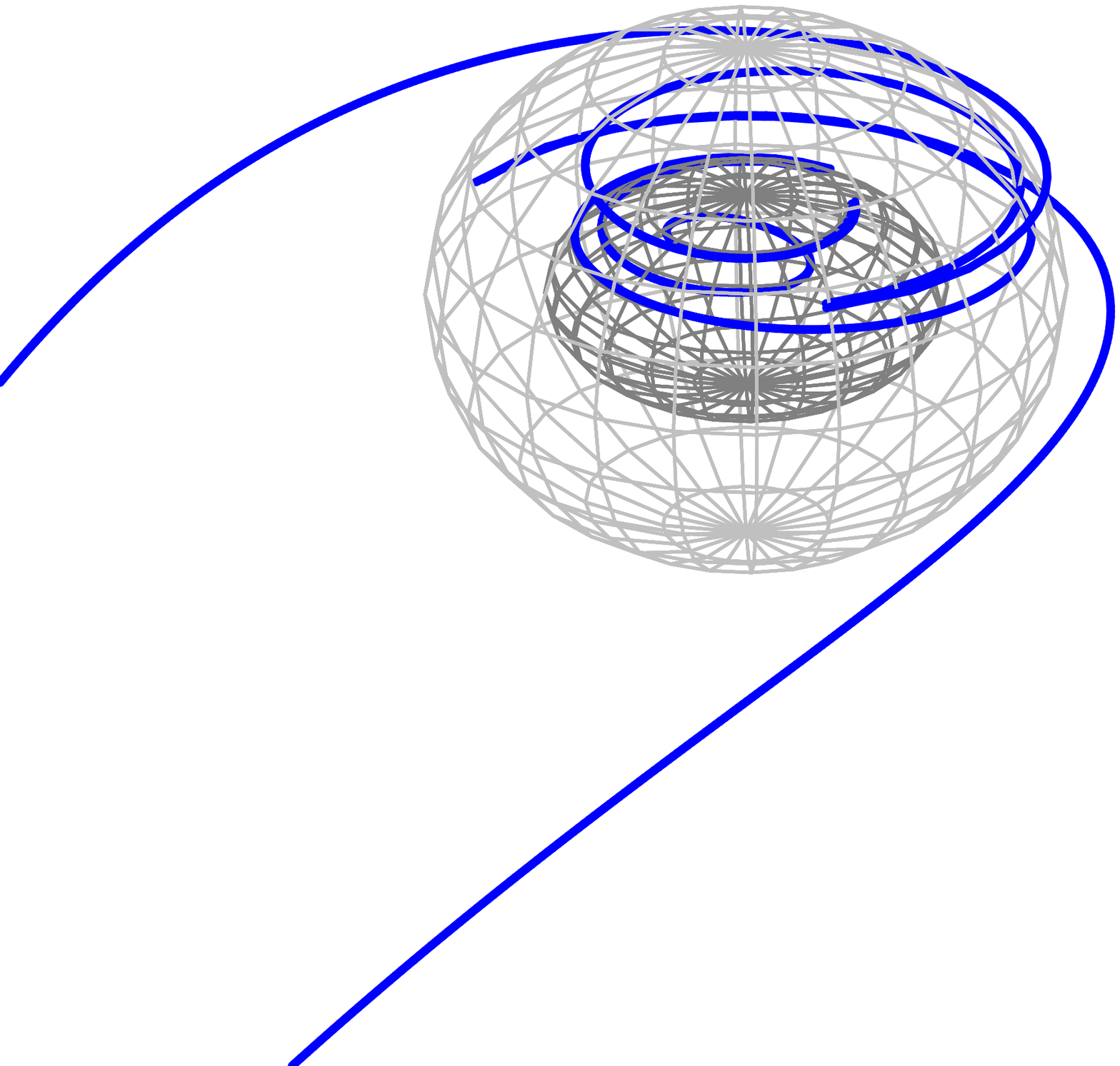}
 }\qquad\qquad
 \subfigure[$x$-$y$-$w$-plot]{
   \includegraphics[width=6cm]{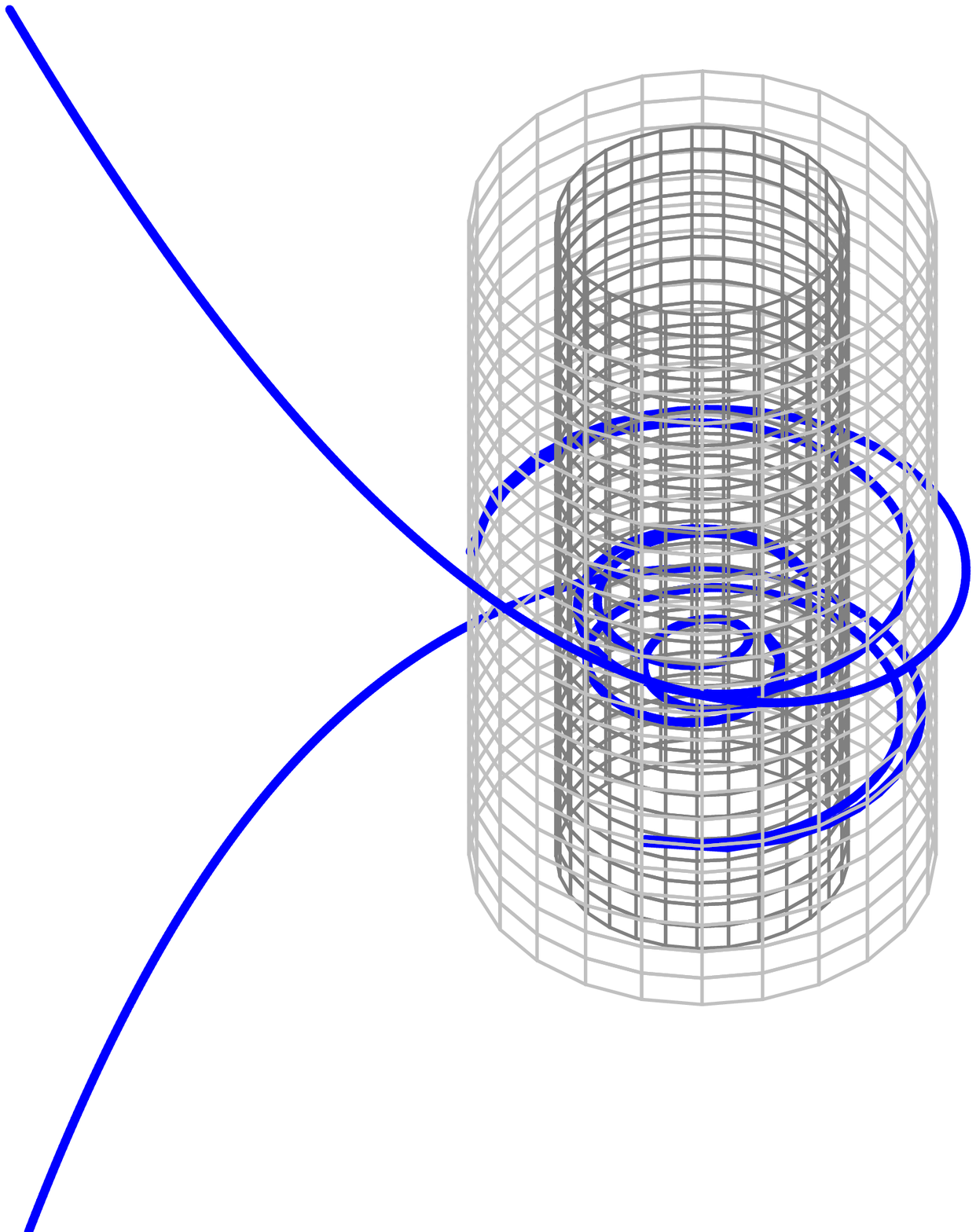}
 }
 \caption{$\delta=0$, $\ta=0.45$, $\tL=-0.5$, $J=2$, $\tK=5$ and $E=2.25$:\newline
          Two-world escape orbit for light in the rotating black string spacetime. The ellipsoids or cylinders are the horizons.}
 \label{pic:kerr-teo-light}
\end{figure}

\begin{figure}[ht]
 \centering
 \subfigure[$x$-$y$-$z$-plot]{
   \includegraphics[width=6cm]{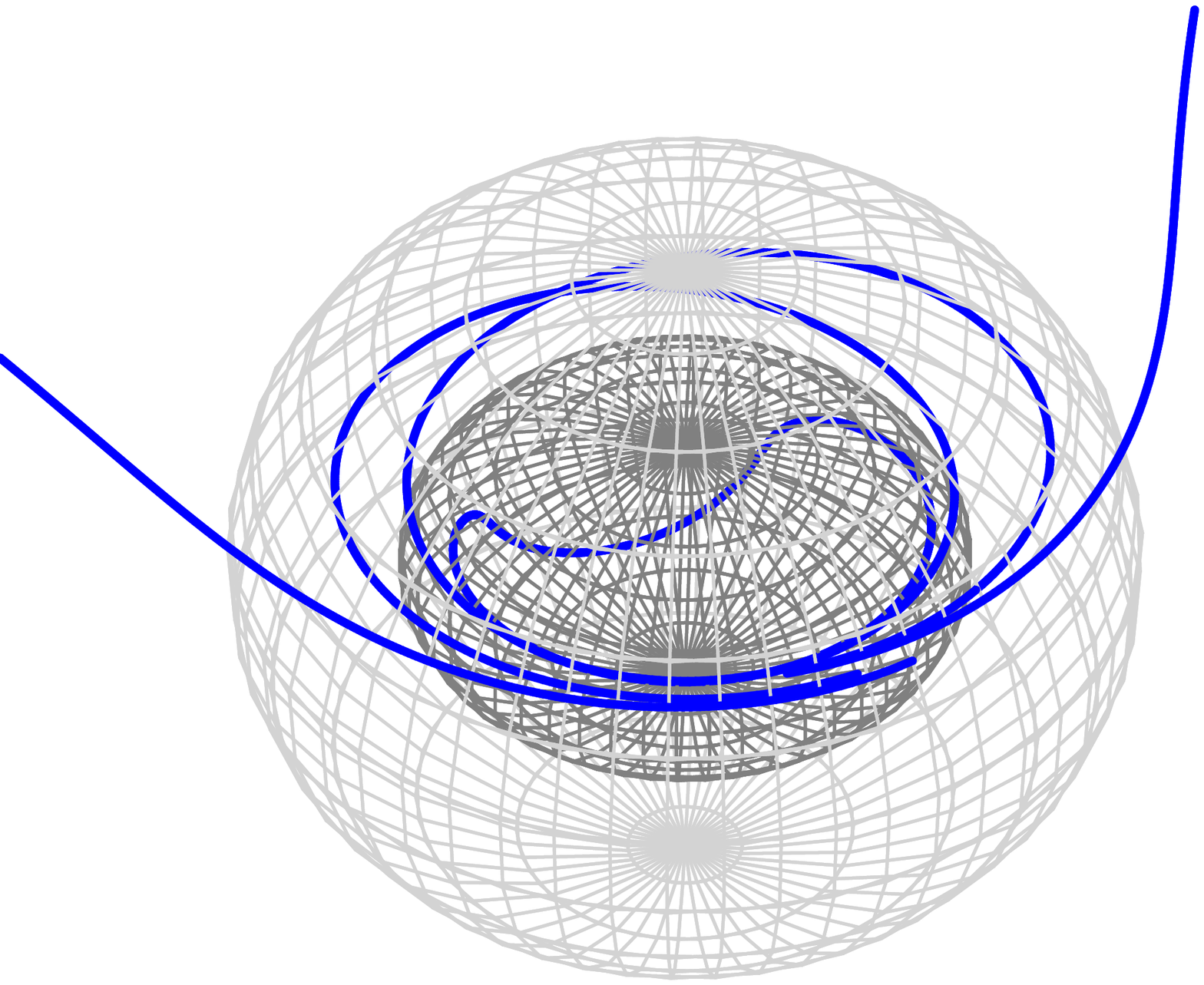}
 }\qquad\qquad
 \subfigure[$x$-$y$-$w$-plot]{
   \includegraphics[width=6cm]{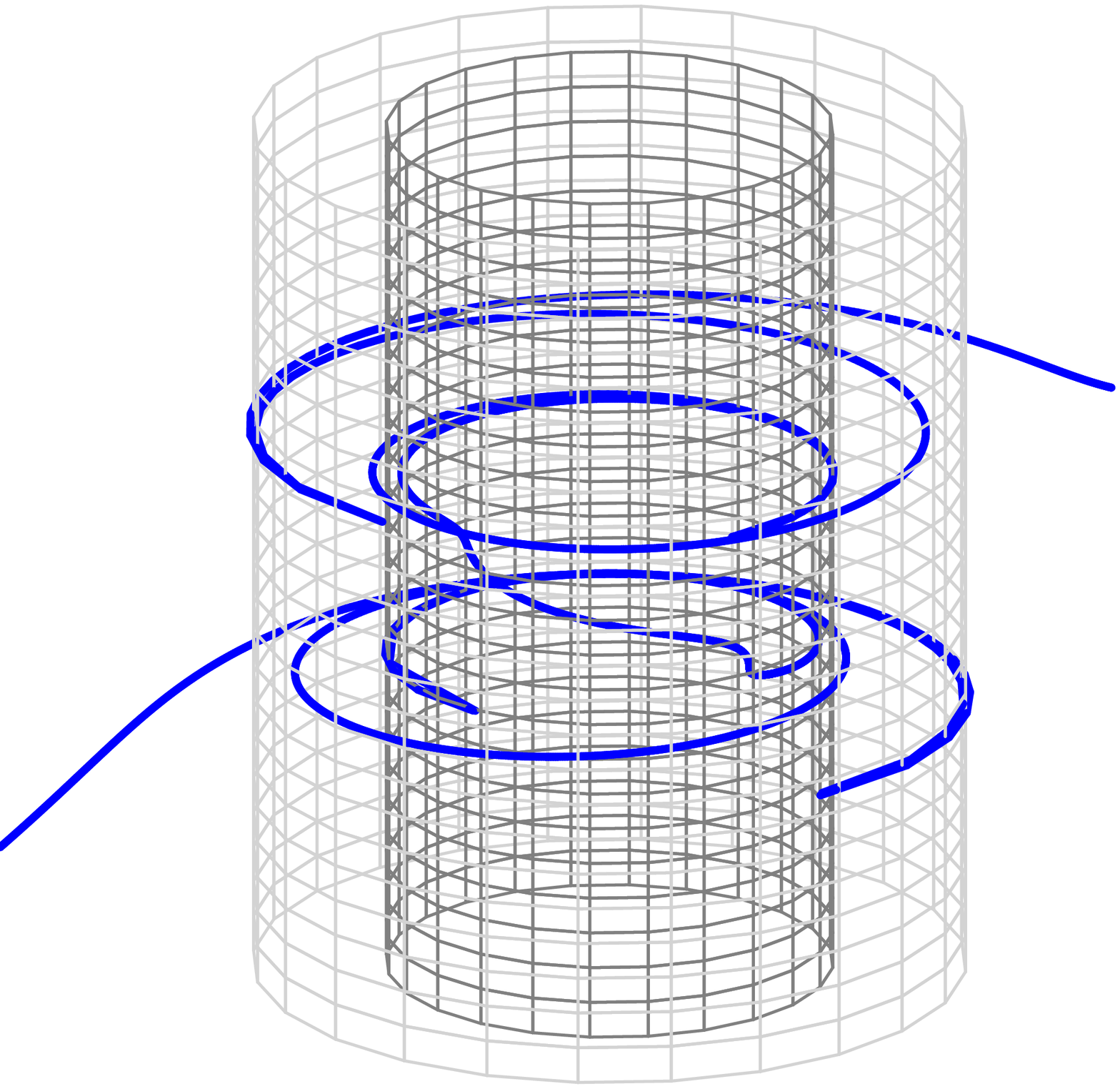}
 }
 \caption{$\delta=1$, $\ta=0.45$, $\tL=0.8$, $J=2$, $\tK=2$ and $E=5.5$:\newline
          Crossover two-world escape orbit for particles in the rotating black string spacetime. The ellipsoids or cylinders are the horizons.}
 \label{pic:kerr-cteo-particle}
\end{figure}

An escape orbit is shown in figure \ref{pic:kerr-eo-particle}. If the turning point of an escape orbit lies behind the two horizons the orbit is called two-world escape orbit (see figure \ref{pic:kerr-teo-light}). Since both horizons are traversed twice, the test particle or light emerges into another universe. It is also possible that the turning point is at negative $\tr$ (see figure \ref{pic:kerr-cteo-particle}), then the orbit is called crossover two-world escape orbit since it does not only traverses both horizons, but also crosses $\tr=0$.

\begin{figure}[ht]
 \centering
 \subfigure[$x$-$y$-$z$-plot]{
   \includegraphics[width=6cm]{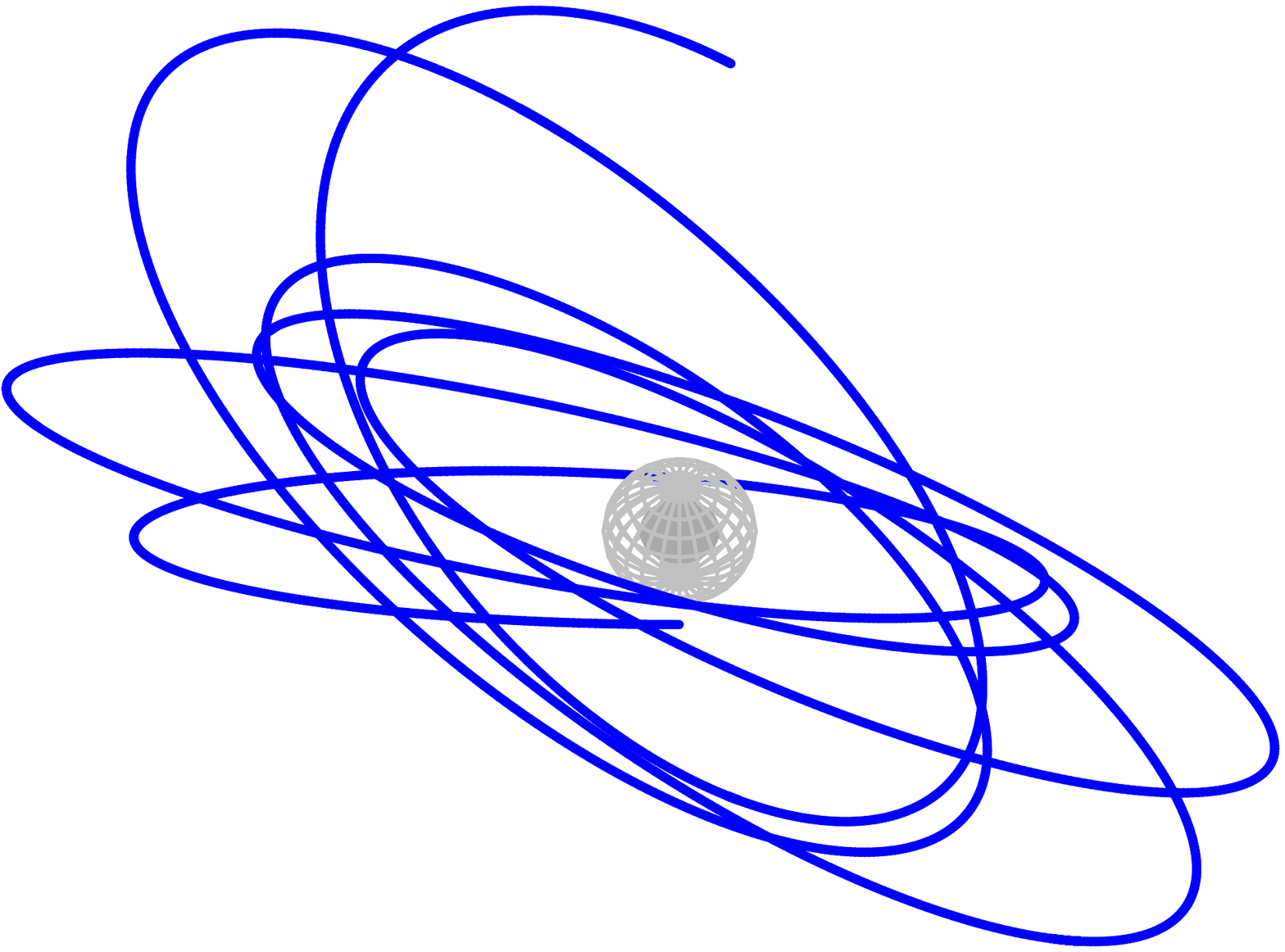}
 }\qquad\qquad
 \subfigure[$x$-$y$-$w$-plot]{
   \includegraphics[width=6cm]{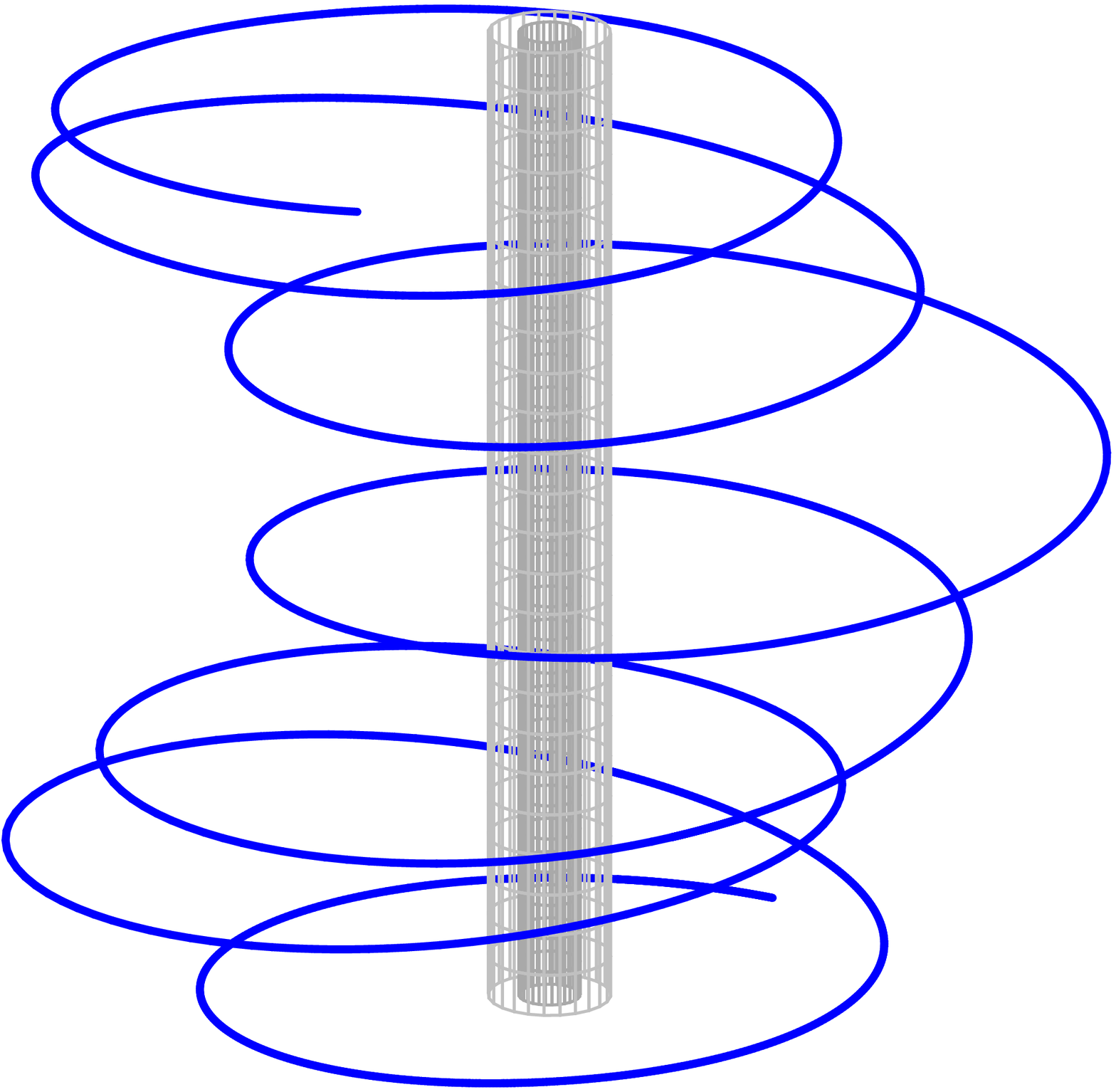}
 }
 \caption{$\delta=0$, $\ta=0.4$, $\tL=2$, $J=\sqrt{2}$, $\tK=5$ and $E=1.36$:\newline
          Bound orbit for light in the rotating black string spacetime. The ellipsoids or cylinders are the horizons.}
 \label{pic:kerr-bo-light}
\end{figure}

A special feature of the the rotating black string spacetime are stable bound orbits of light (see figure \ref{pic:kerr-bo-light}). Such orbits do not exist in the ordinary four-dimensional Kerr spacetime.

\begin{figure}[ht]
 \centering
 \subfigure[$x$-$y$-$z$-plot]{
   \includegraphics[width=6cm]{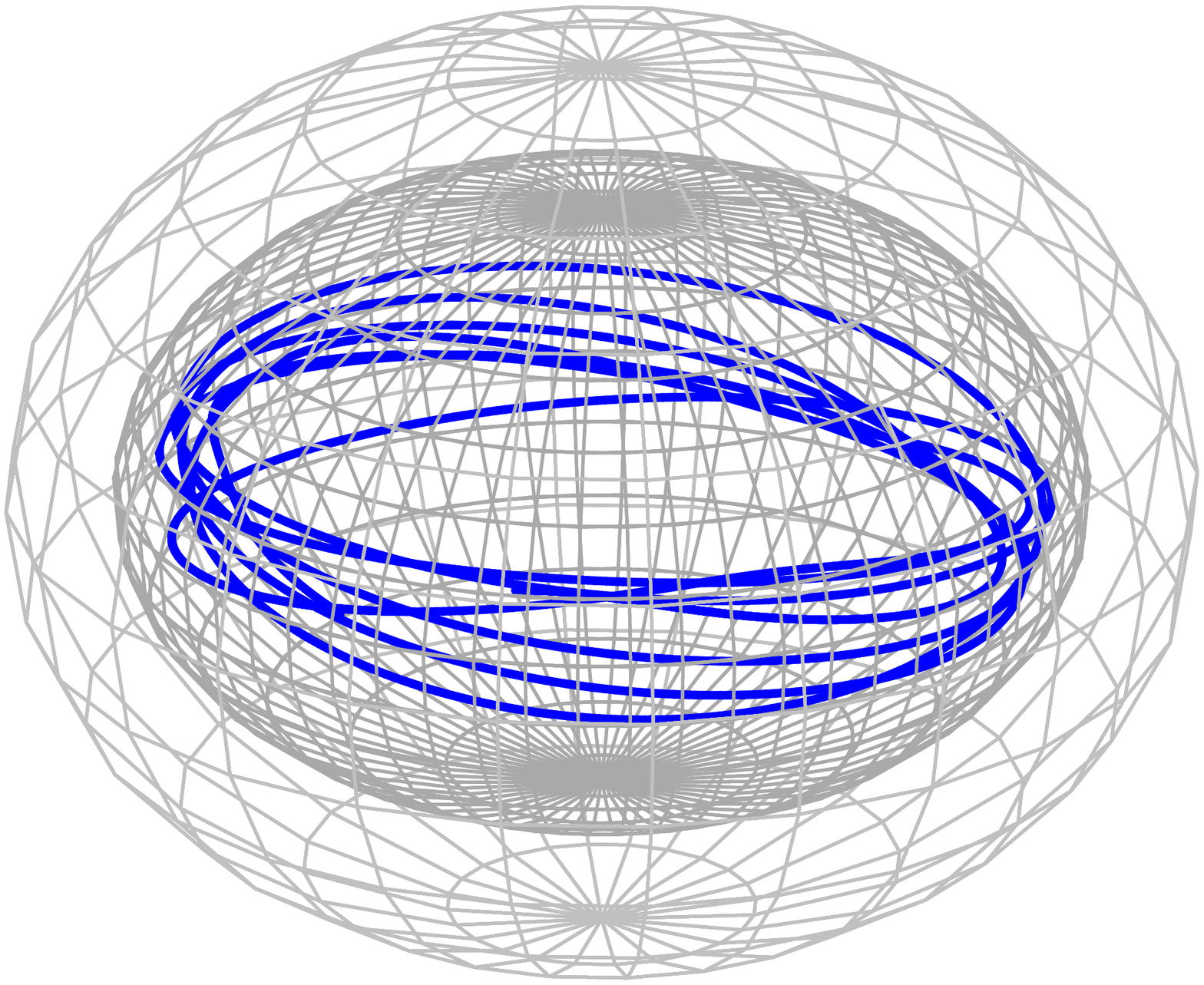}
 }\qquad\qquad
 \subfigure[$x$-$y$-$w$-plot]{
   \includegraphics[width=6cm]{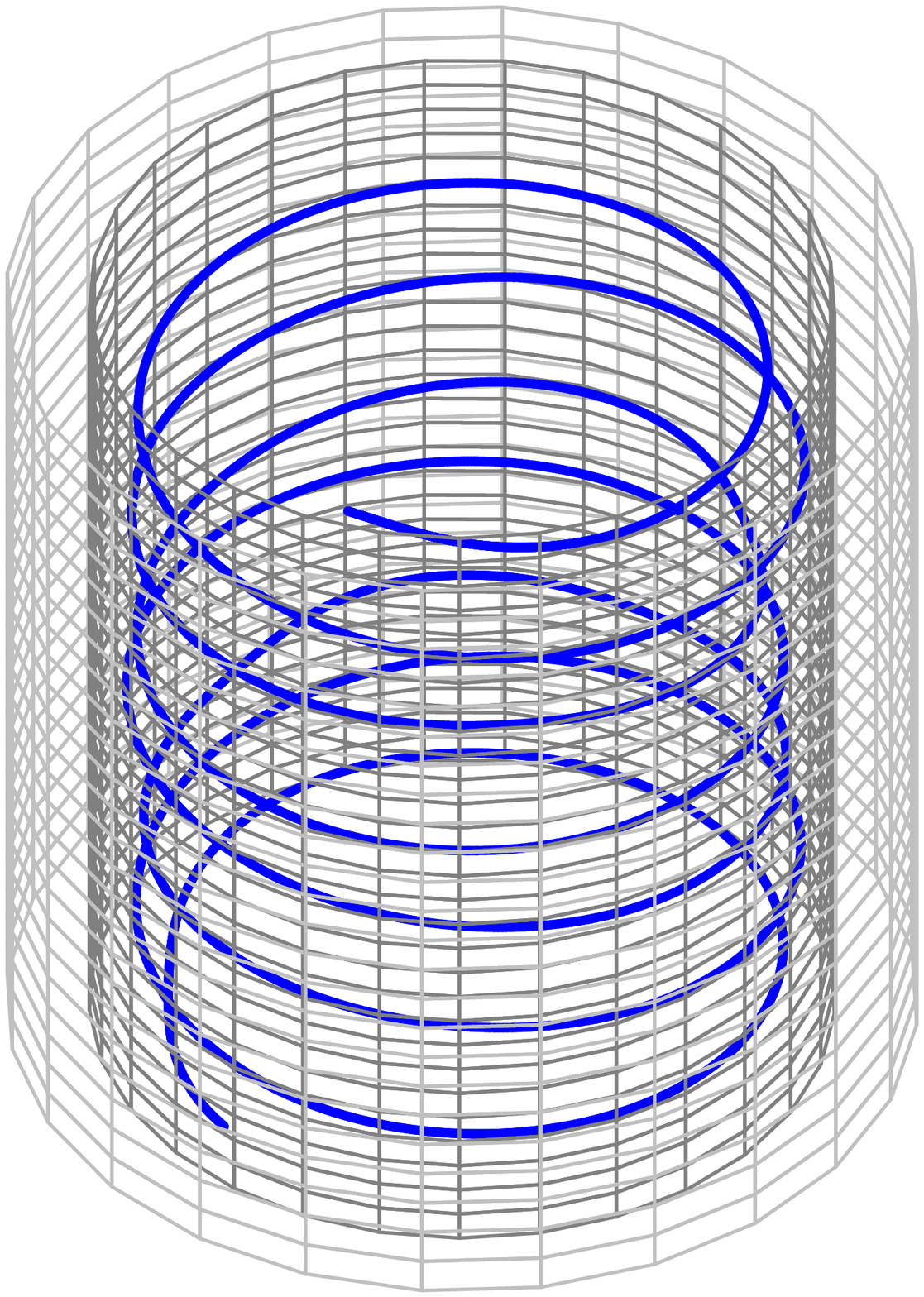}
 }
 \caption{$\delta=1$, $\ta=0.49$, $\tL=1.4$, $J=0.5$, $\tK=0.6$ and $E=1.63$:\newline
          Bound orbit for particles hidden behind the inner horizon in the rotating black string spacetime. The ellipsoids or cylinders are the horizons.}
 \label{pic:kerr-innerbo-particle}
\end{figure}

Like in the Kerr spacetime there are also bound orbits hidden behind the inner horizon $\tr_-$ if the Carter constant $\tK$ is rather small (see figure \ref{pic:kerr-innerbo-particle}).

\begin{figure}[ht]
 \centering
 \subfigure[$x$-$y$-$z$-plot]{
   \includegraphics[width=6cm]{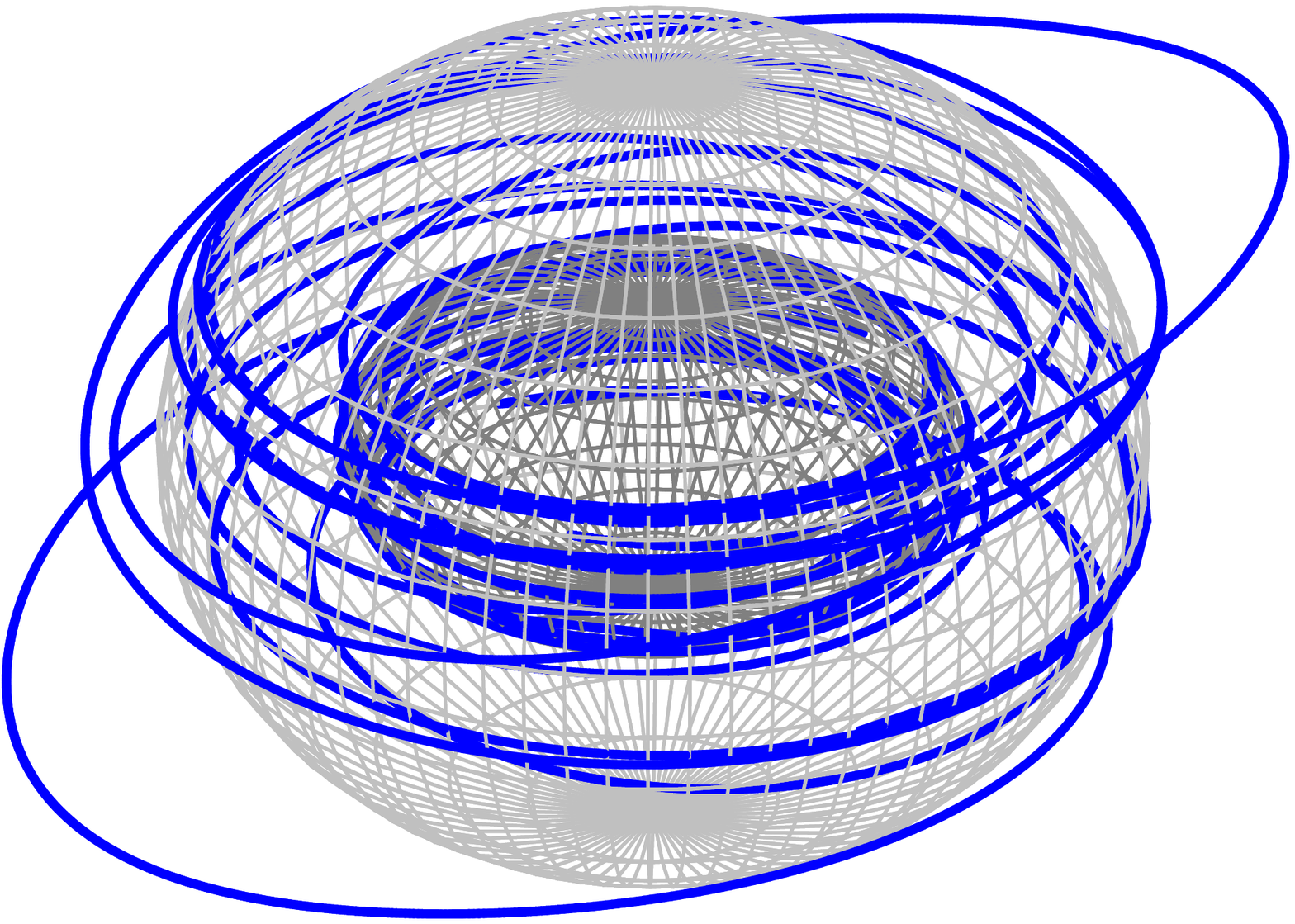}
 }\qquad\qquad
 \subfigure[$x$-$y$-$w$-plot]{
   \includegraphics[width=6cm]{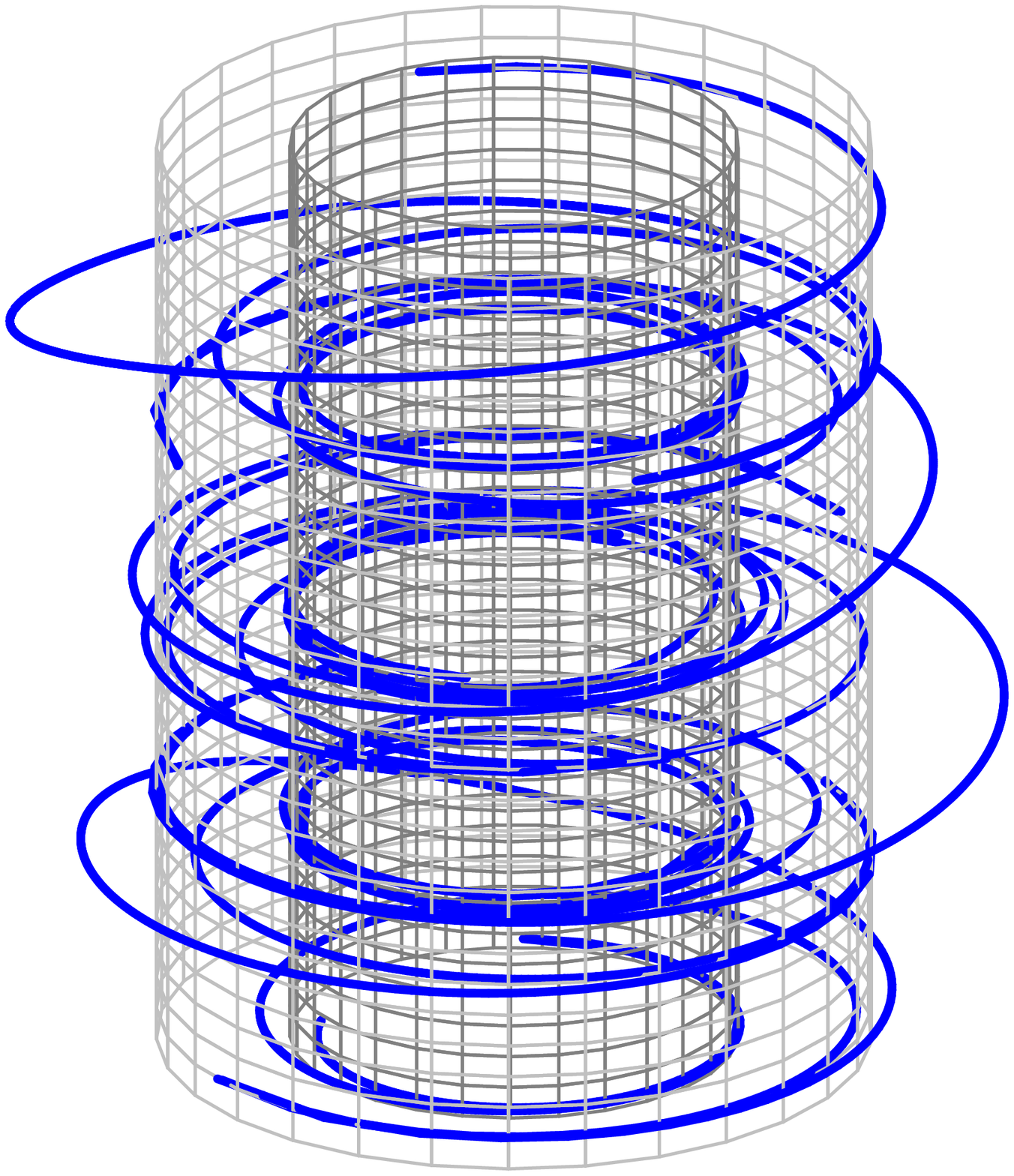}
 }
 \caption{$\delta=1$, $\ta=0.45$, $\tL=0.1$, $J=5$, $\tK=1.8$ and $E=2.4$:\newline
          Many-world bound orbit for particles in the rotating black string spacetime. The ellipsoids or cylinders are the horizons.}
 \label{pic:kerr-mbo-particle}
\end{figure}

It is also possible that a particle or light on a bound orbit crosses both horizons several times. Each time both horizons are traversed twice, the orbit continues in another universe. Such a many-world bound orbit is depicted in figure \ref{pic:kerr-mbo-particle}.

\begin{figure}[ht]
 \centering
 \subfigure[$x$-$y$-$z$-plot]{
   \includegraphics[width=6.5cm]{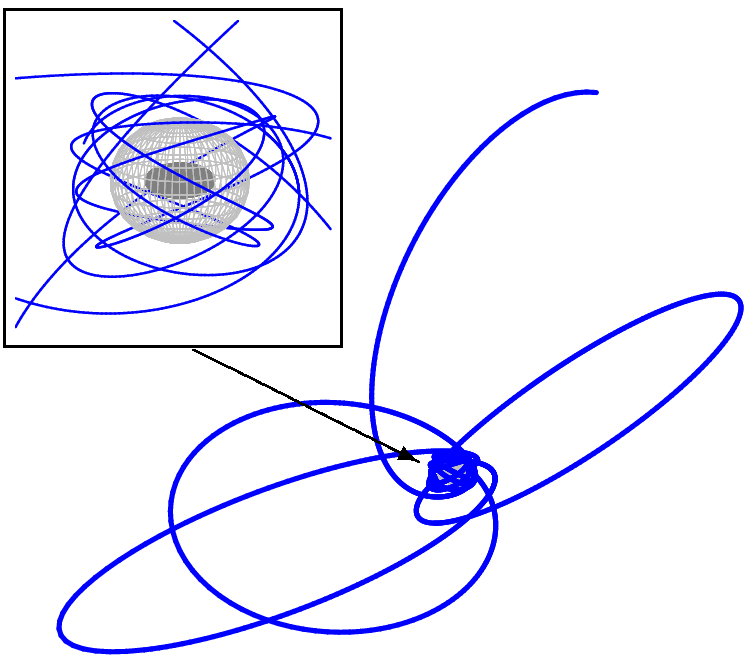}
 }\qquad\qquad
 \subfigure[$x$-$y$-$w$-plot]{
   \includegraphics[width=6.5cm]{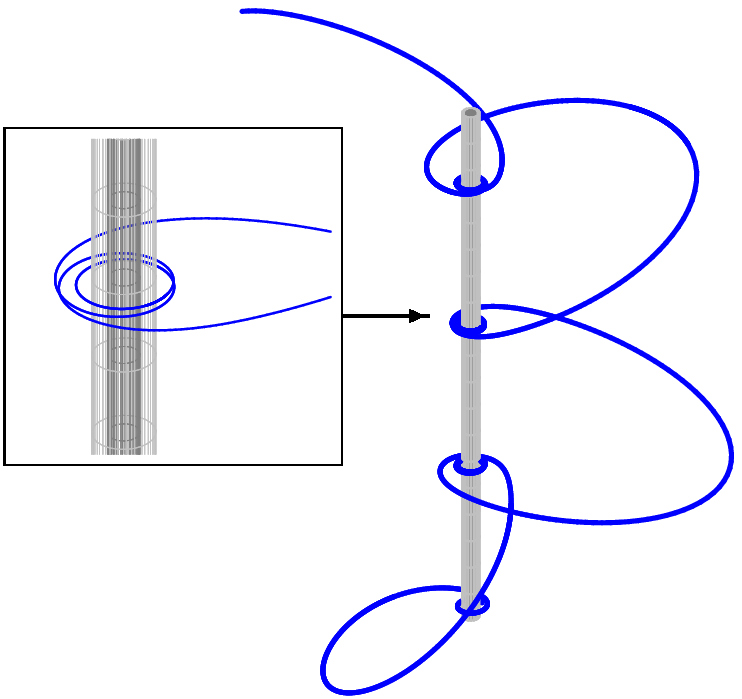}
 }
 \caption{$\delta=1$, $\ta=0.4$, $\tL=2$, $J=\sqrt{2}$, $\tK=5$ and $E=1.706$:\newline
          ``Zoom-whirl'' bound orbit for particles in the rotating black string spacetime. The ellipsoids or cylinders are the horizons.}
 \label{pic:zoomwhirl-bo-particle}
\end{figure}

Figure \ref{pic:zoomwhirl-bo-particle} shows a so called ``zoom-whirl'' bound orbit (see also \cite{Schmidt:2002qk,Glampedakis:2002ya,Levin:2008yp}). Here the energy of the test particle is very close to a local maximum of the effective potential, so that the particle ``whirls'' around the black string before reaching the periastron and then ``zooms'' out to return to its elliptical orbit where it moves to the apastron.

\clearpage

\section{Conclusion}

In this article we presented the complete set of analytical solutions of the geodesic equations of test particles and light in the static and the rotating black string spacetime. The static and the rotating black string metric are obtained by adding a compact dimension to the Schwarzschild and Kerr metric.

The geodesic equations can be solved in terms of the elliptic $\wp$-, $\sigma$- and $\zeta$-functions. Using effective potential techniques and parametric diagrams, possible types of orbits were derived. In the static case BOs, EOs and TOs are possible, while in the rotating case BOs, MBOs, EOs, TEOs, CTEOs, TrOs and TOs are possible.

In contrast to the ordinary four-dimensional Schwarzschild and Kerr spacetime, bound orbits of light are possible both in the static and the rotating black string spacetime.\\

Hereby the analytic solutions do not only give a proof of the existence of bound orbits of light in the black string spacetime. They also present a usefull tool to calculate the exact orbits and their properties, including observables like the periastron shift of bound orbits, the light deflection of escape orbits, the deflecton angle and the Lense-Thirring effect. For the calculation of the observables analogous formulas to those given in \cite{Hackmann:2010zz} can be used. Observables can later be compared to observations.\\

The black strings considered here are so-called uniform black strings, since there is no dependence of the extra dimension. However, uniform black strings are subject to the Gregory-Laflamme instability \cite{Gregory:1993vy}, which is associated with the emergence of a branch of non-uniform black strings, whose horizon size is not constant w.r.t.~the compact direction, but depends on the compact coordinate \cite{Gubser:2001ac,Wiseman:2002zc,Kleihaus:2006ee}. At the endpoint of this nonuniform black string branch a horizon topology changing transition should be encountered \cite{Kol:2002xz,Kol:2004ww}.

To gain a better understanding of this transition, it is essential to solve the geodesic equations. However, for the construction of nonuniform black strings no analytic techniques are available, and numerical techniques have to be employed. Likewise, the geodesic equations must be solved numerically. It is therefore of high relevance, to have analytic solutions of the geodesic equations available as a testbed. Moreover, it will be very interesting to see, how the set of analytic solutions changes, as the black strings become more and more nonuniform.

Also non-uniform rotating and charged black strings have been considered in \cite{Kleihaus:2007dg} and \cite{Kleihaus:2009ff} whose geodesics would be very interesting to study.\\

It is also interesting to consider higher dimensional ($d>4$) Schwarzschild spacetimes, here stable bound orbits are no longer possible \cite{Hackmann:2008tu}. One may wonder wether bound orbits of light become possible if a compact dimension is added to the metric. But it can be shown that unlike in the (rotating) black string spacetime in five dimensions, stable bound orbits of light are not possible if a compact dimension is added to higher dimensional Schwarzschild spacetimes.\\

Moreover the bound orbits of light appear neither in the spacetime of an Abelian-Higgs string \cite{Hartmann:2010rr} nor in the spacetime of two interacting Abelian-Higgs strings \cite{Hartmann:2012pj}. There escape orbits are the only possibility for massless particles. For cosmic superstrings bound orbits of light are possible in Melvin spacetimes but not in asymptotically conical spacetimes \cite{Hartmann:2010vp}.

\section{Acknowledgements}

We gratefully acknowledge support by the DAAD and the DFG, in particular, within the DFG Research Training Group 1620 ``Models of Gravity''. Also we would like to thank Jutta Kunz, Valeria Kagramanova and Burkhard Kleihaus for helpful discussions. B.K. would like to thank Suneeta Vardarajan for her guidance.


\bibliographystyle{unsrt}

\begin{thebibliography}{99}

\bibitem{Schwarzschild:1916ae} 
  K.~Schwarzschild,
  Sitzungsber.\ Preuss.\ Akad.\ Wiss.\ Berlin (Math.\ Phys.\ ) {\bf 1916}, 424 (1916)
  [physics/9912033].

\bibitem{Kerr:1963ud} 
  R.~P.~Kerr,
  Phys.\ Rev.\ Lett.\  {\bf 11}, 237 (1963).

\bibitem{Kaluza:1921tu} 
  T.~Kaluza,
  Sitzungsber.\ Preuss.\ Akad.\ Wiss.\ Berlin (Math.\ Phys.\ ) {\bf 1921}, 966 (1921).

\bibitem{Klein:1926tv} 
  O.~Klein,
  Z.\ Phys.\  {\bf 37}, 895 (1926)
  [Surveys High Energ.\ Phys.\  {\bf 5}, 241 (1986)].

\bibitem{Hagihara:1931}
 Y.~Hagihara,
 Jpn.\ J.\ Astron.\ Geophys.\ {\bf 8}, 67 (1931)

\bibitem{Kagramanova:2010bk} 
  V.~Kagramanova, J.~Kunz, E.~Hackmann and C.~L\"ammerzahl,
  Phys.\ Rev.\ D {\bf 81}, 124044 (2010)
  [arXiv:1002.4342 [gr-qc]].

\bibitem{Grunau:2010gd} 
  S.~Grunau and V.~Kagramanova,
  Phys.\ Rev.\ D {\bf 83}, 044009 (2011)
  [arXiv:1011.5399 [gr-qc]].

\bibitem{Kagramanova:2012hw} 
  V.~Kagramanova and S.~Reimers,
  Phys.\ Rev.\ D {\bf 86}, 084029 (2012)
  [arXiv:1208.3686 [gr-qc]].

\bibitem{Hackmann:2008zza} 
  E.~Hackmann and C.~L\"ammerzahl,
  Phys.\ Rev.\ Lett.\  {\bf 100}, 171101 (2008).

\bibitem{Hackmann:2008zz} 
  E.~Hackmann and C.~L\"ammerzahl,
  Phys.\ Rev.\ D {\bf 78}, 024035 (2008).

\bibitem{Hackmann:2008tu} 
  E.~Hackmann, V.~Kagramanova, J.~Kunz and C.~L\"ammerzahl,
  Phys.\ Rev.\ D {\bf 78}, 124018 (2008)
  [Erratum-ibid.\  {\bf 79}, 029901 (2009)]
  [arXiv:0812.2428 [gr-qc]].

\bibitem{Hackmann:2010zz} 
  E.~Hackmann, C.~L\"ammerzahl, V.~Kagramanova and J.~Kunz,
  Phys.\ Rev.\ D {\bf 81}, 044020 (2010)
  [arXiv:1009.6117 [gr-qc]].

\bibitem{Enolski:2010if} 
  V.~Z.~Enolski, E.~Hackmann, V.~Kagramanova, J.~Kunz and C.~L\"ammerzahl,
  J.\ Geom.\ Phys.\  {\bf 61}, 899 (2011)
  [arXiv:1011.6459 [gr-qc]].

\bibitem{Grunau:2012ai} 
  S.~Grunau, V.~Kagramanova, J.~Kunz and C.~L\"ammerzahl,
  Phys.\ Rev.\ D {\bf 86}, 104002 (2012)
  [arXiv:1208.2548 [gr-qc]].

\bibitem{Grunau:2012ri} 
  S.~Grunau, V.~Kagramanova and J.~Kunz,
  Phys.\ Rev.\ D {\bf 87}, 044054 (2013)
  arXiv:1212.0416 [gr-qc].

\bibitem{Aliev:1988wv} 
  A.~N.~Aliev and D.~V.~Galtsov,
  Sov.\ Astron.\ Lett.\  {\bf 14}, 48 (1988).

\bibitem{Galtsov:1989ct} 
  D.~V.~Galtsov and E.~Masar,
  Class.\ Quant.\ Grav.\  {\bf 6}, 1313 (1989).

\bibitem{Chakraborty:1991mb} 
  S.~Chakraborty and L.~Biswas,
  Class.\ Quant.\ Grav.\  {\bf 13}, 2153 (1996).

\bibitem{Ozdemir:2003km} 
  N.~Ozdemir,
  Class.\ Quant.\ Grav.\  {\bf 20}, 4409 (2003).

\bibitem{Ozdemir:2004ne} 
  F.~Ozdemir, N.~Ozdemir and B.~T.~Kaynak,
  Int.\ J.\ Mod.\ Phys.\ A {\bf 19}, 1549 (2004).

\bibitem{Hackmann:2009rp} 
  E.~Hackmann, B.~Hartmann, C.~L\"ammerzahl and P.~Sirimachan,
  Phys.\ Rev.\ D {\bf 81}, 064016 (2010)
  [arXiv:0912.2327 [gr-qc]].

\bibitem{Hackmann:2010ir} 
  E.~Hackmann, B.~Hartmann, C.~L\"ammerzahl and P.~Sirimachan,
  Phys.\ Rev.\ D {\bf 82}, 044024 (2010)
  [arXiv:1006.1761 [gr-qc]].

\bibitem{Hartmann:2010rr} 
  B.~Hartmann and P.~Sirimachan,
  JHEP {\bf 1008}, 110 (2010)
  [arXiv:1007.0863 [gr-qc]].

\bibitem{Hartmann:2012pj} 
  B.~Hartmann and V.~Kagramanova,
  Phys.\ Rev.\ D {\bf 86}, 045028 (2012)
  [arXiv:1204.0396 [hep-th]].

\bibitem{Hartmann:2010vp} 
  B.~Hartmann, C.~L\"ammerzahl and P.~Sirimachan,
  Phys.\ Rev.\ D {\bf 83}, 045027 (2011)
  [arXiv:1012.3285 [hep-th]].

\bibitem{Mino:2003yg} 
  Y.~Mino,
  Phys.\ Rev.\ D {\bf 67}, 084027 (2003)
  [gr-qc/0302075].

\bibitem{Markushevich:1967}
  A.~I.~Markushevich, {\em Theory of Functions of a Complex Variable} (Prentice-Hall, Englewood Cliffs, NJ, 1967), Vol. III.

\bibitem{ONeill:1995}
 B.~O'Neill, {\em The Geometry of Kerr Black Holes} (AK Peters, Wellesley, Massachusetts, 1995).

\bibitem{Carter:1968rr} 
  B.~Carter,
  Phys.\ Rev.\  {\bf 174}, 1559 (1968).

\bibitem{Carter:1966zza} 
  B.~Carter,
  Phys.\ Rev.\  {\bf 141}, 1242 (1966).

\bibitem{Schmidt:2002qk} 
  W.~Schmidt,
  Class.\ Quant.\ Grav.\  {\bf 19}, 2743 (2002)
  [gr-qc/0202090].

\bibitem{Glampedakis:2002ya} 
  K.~Glampedakis and D.~Kennefick,
  Phys.\ Rev.\ D {\bf 66}, 044002 (2002)
  [gr-qc/0203086].

\bibitem{Levin:2008yp} 
  J.~Levin and G.~Perez-Giz,
  Phys.\ Rev.\ D {\bf 79}, 124013 (2009)
  [arXiv:0811.3814 [gr-qc]].

\bibitem{Gregory:1993vy} 
  R.~Gregory and R.~Laflamme,
  Phys.\ Rev.\ Lett.\  {\bf 70}, 2837 (1993)
  [hep-th/9301052].

\bibitem{Gubser:2001ac} 
  S.~S.~Gubser,
  Class.\ Quant.\ Grav.\  {\bf 19}, 4825 (2002)
  [hep-th/0110193].

\bibitem{Wiseman:2002zc} 
  T.~Wiseman,
  Class.\ Quant.\ Grav.\  {\bf 20}, 1137 (2003)
  [hep-th/0209051].

\bibitem{Kleihaus:2006ee} 
  B.~Kleihaus, J.~Kunz and E.~Radu,
  JHEP {\bf 0606}, 016 (2006)
  [hep-th/0603119].

\bibitem{Kol:2002xz} 
  B.~Kol,
  JHEP {\bf 0510}, 049 (2005)
  [hep-th/0206220].

\bibitem{Kol:2004ww} 
  B.~Kol,
  Phys.\ Rept.\  {\bf 422}, 119 (2006)
  [hep-th/0411240].

\bibitem{Kleihaus:2007dg} 
  B.~Kleihaus, J.~Kunz and E.~Radu,
  JHEP {\bf 0705}, 058 (2007)
  [hep-th/0702053]

\bibitem{Kleihaus:2009ff} 
  B.~Kleihaus, J.~Kunz, E.~Radu and C.~Stelea,
  JHEP {\bf 0909}, 025 (2009)
  [arXiv:0905.4716 [hep-th]].

\end{thebibliography}

\end{document}